\documentclass[prd,showpacs,amsmath,amssymb]{revtex4}
\usepackage{graphicx,color}

\begin{document}

\title{New Hadron States}

\author{Shi-Lin Zhu}\email{zhusl@phy.pku.edu.cn }
\affiliation{Department of Physics, Peking University, Beijing
100871, China}

\vspace*{1.0cm}

\date{\today}

\begin{abstract}
The past four years has witnessed the renaissance of the hadron
spectroscopy. Many interesting new hadron states were discovered
experimentally, some of which do not fit into the quark model
easily. I will give a concise overview of these states and their
possible interpretations. Topics covered in this review are: (1)
candidates of new light hadrons including $p {\bar p}$ threshold
enhancement, X(1835), X(1576), $f_0(1810)$, recent candidates of
the $1^{-+}$ exotic mesons, Y(2175), $p\bar\Lambda$ threshold
enhancement etc; (2) charmed mesons including p-wave non-strange
charmed mesons, $D_{sj}(2317)$ and $D_{sj}(2460)$, recent
candidates of higher excited charmed mesons, $D_{sj}(2632)$ etc;
(3) charmonium and charmonium-like states such as X(3872),
Y(4260), X(3940), Y(3940), Z(3930) etc. The effect from the nearby
S-wave open channels on the quark model spectrum above or near
strong decay threshold is emphasized. Dynamical lattice
simulations of $D K$ and $D^0\bar D^{\ast 0}$ scattering and the
extraction of their phase shifts may help resolve the underlying
structure of $D_{sj}(2317)$, $D_{sj}(2460)$ and X(3872).
\end{abstract}


\maketitle

\newpage

\begin{center}
{\bf\Large {Table of contents}}
\end{center}
\vspace{0.5cm}

\begin{itemize}

\item 1. Introduction

\item 2. Candidates of new light hadrons

              \begin{itemize}

              \item  $p {\bar p}$ threshold enhancement {\sl vs} X(1835)

              \item  X(1576)

              \item  $f_0(1810)$

              \item  Recent candidates of the $1^{-+}$ exotic mesons

              \item  Y(2175)

              \item  $p\bar\Lambda$ threshold enhancement

              \end{itemize}

\item 3. Charmed hadrons

           \begin{itemize}

              \item P-wave non-strange charmed mesons

              \item $D_{sj}(2317)$ and $D_{sj}(2460)$

                      \begin{itemize}
                      \item Interpretations of $D_{sj}(2317)$ and $D_{sj}(2460)$
                      \item Radiative decays of $D_{sJ}(2317)$ and $D_{sJ}(2460)$
                      \item Strong decays of $D_{sJ}(2317)$ and $D_{sJ}(2460)$
                      \item The low mass puzzle of $D_{sj}(2317)$ and $D_{sj}(2460)$
                      \end{itemize}

              \item Recent candidates of higher excited charmed mesons

              \item $D_{sj}(2632)$

              \end{itemize}

\item 4. Charmonium or charmonium-like states

           \begin{itemize}

              \item X(3872)
                      \begin{itemize}
                      \item The discovery of X(3872)
                      \item Theoretical interpretations of X(3872)
                      \item Experimental evidence against the molecular assignment
                      \item Is X(3872) still a $1^{++}$ charmonium?
                      \end{itemize}

              \item Y(4260), Y(4385)

              \item X(3940), Y(3940), Z(3930)

              \end{itemize}

\item 5. Conclusions

\end{itemize}

\newpage

%%%%%%%%%%%%%%%%%%%%%%%%%%%%%%%%%%%
\section{Introduction}\label{sec1}
%%%%%%%%%%%%%%%%%%%%%%%%%%%%%%%%%%%

According to the constituent quark model (CQM), mesons and baryons
are composed of $q\bar q$ and $qqq$ respectively. CQM provides a
convenient framework in the classification of hadrons. Most of
experimentally observed hadron states fit into this scheme quite
neatly. Any states beyond CQM are labelled as "non-conventional"
hadrons.

However, CQM is only a phenomenological model. It's not derived
from the underlying theory of the strong interaction---Quantum
Chromodynamics (QCD). Hence the CQM spectrum is not necessarily
the same as the physical spectrum of QCD, which remains ambiguous
and elusive after decades of intensive theoretical and
experimental exploration. No one is able to either prove or
exclude the existence of these non-conventional states rigorously
since the QCD confinement issue is not solved yet.

Hadron physicists generally take a modest and practical attitude.
Suppose these non-conventional states exist. Then the important
issues are: (1) How to determine their characteristic quantum
numbers and estimate their masses, production cross-section and
decay widths reliably? (2) How and in which channels to dig out
the signal from backgrounds and identify them experimentally?

There are three classes of "non-conventional" hadrons. The first
class are mesons with "exotic" $J^{PC}$ quantum numbers. The
possible angular momentum, parity and charge conjugation parity of
a neutral $q\bar q $ meson are $J^{PC}=0^{++}, 0^{-+}, 1^{++},
1^{--}, 1^{+-}, \cdots$. In other words, a $q\bar q $ meson can
never have $J^{PC}=0^{--}, 0^{+-}, 1^{-+}, 2^{+-}, 3^{-+}, \cdots
$. Any state with these "exotic" quantum numbers is clearly beyond
CQM. We want to emphasize they are "exotic" only in the context of
CQM. One can construct color-singlet local operators to verify
that these quantum numbers are allowed in QCD. "Exotic" quantum
numbers provide a powerful method for the experimental search of
these "non-conventional" states. In contrast, a $qqq$ baryon in
CQM exhausts all possible $J^P$, i.e., $J^P={1\over 2}^{\pm},
{3\over 2}^{\pm}, {5\over 2}^{\pm}, \cdots$.

The second class are hadrons with exotic flavor content. One
typical example is the $\Theta^+$ pentaquark. It was discovered in
$K^+ n$ channel with the quark content $uudd\bar s$ \cite{nakano}.
Such a state is clearly beyond CQM. Exotic flavor content is also
an asset in the experimental search of them.

The third class are hadrons which have ordinary quantum numbers
but do not easily fit into CQM. Let's take the $J^{PC}=0^{++}$
scalar mesons as an example. Below 2 GeV, we have $\sigma,
f_0(980), f_0(1370), f_0(1500), f_0(1710), f_0(1790), f_0(1810)$.
Within CQM there are only two scalars within this mass range if we
ignore the radial excitations. With radial excitations, CQM could
accommodate four scalars at most. Clearly there is serious
overpopulation of the scalar spectrum. If all the above states are
genuine, the quark content of some of them is not $q\bar q$.
Overpopulation of the spectrum provides another useful window in
the experimental search of non-conventional states.

Glueballs are hadrons composed of gluons. Quenched lattice QCD
simulation suggests the scalar glueball is the lightest. Its mass
is around $(1.5\sim 1.7)$ GeV \cite{lattice}. Glueballs with the
other quantum numbers are high-lying. In the large $N_c$ limit,
glueballs decouple from the conventional $q\bar q$ mesons
\cite{witten}. Moreover, one gluon splits into two gluons freely
in this limit. Hence the number of gluon field inside glueballs is
indefinite when $N_c\to \infty$. In the real world with $N_c=3$,
glueballs mix with nearby $q\bar q$ mesons with the same quantum
numbers, which renders the experimental identification of scalar
glueballs very difficult.

In this review I discuss only new hadrons or candidates of new
hadrons observed in the past four years, especially those which do
not fit into the quark model framework easily. I will not touch
upon bayons including the $\Theta^+$ pentaquark. Experimental
progress in the field of excited charmed baryons and light baryons
is beyond the present short review. Now experiments reporting
negative results are so overwhelming that $\Theta^+$ is probably
an experimental artifact. Interested readers may consult
specialized reviews in Refs. \cite{zhu0,oka0,hicks}. There has
been important experimental progress in the sector of low-lying
scalar mesons below 1 GeV in the past several years while
theoretical interpretations of these states remain very
controversial. This topic deserves a separate review. Hence I also
leave it out completely.

This paper is organized as follows. The first section is devoted
to candidates of new light hadrons. Section \ref{sec3} is on
charmed mesons while Section \ref{sec4} discusses the charmonium
or charmonium-like states. The last section is a short summary.

%%%%%%%%%%%%%%%%%%%%%%%%%%%%%%%%%%%%%%%%%%%%%%%%%
\section{Candidates of new light hadrons}\label{sec2}
%%%%%%%%%%%%%%%%%%%%%%%%%%%%%%%%%%%%%%%%%%%%%%%%%

In the past several years, there has been important progress in
the sector of light hadrons. Quite a few of them were observed
near threshold.

%%%%%%%%%%%%%%%%%%%%%%%%%%%%%%%%%%%%%%%%%%%%%%%%%%%%%%%%%%%%%%%
\subsection{$p {\bar p}$ threshold enhancement {\sl vs} X(1835)}

\begin{figure}[th]
\scalebox{0.5}{\includegraphics{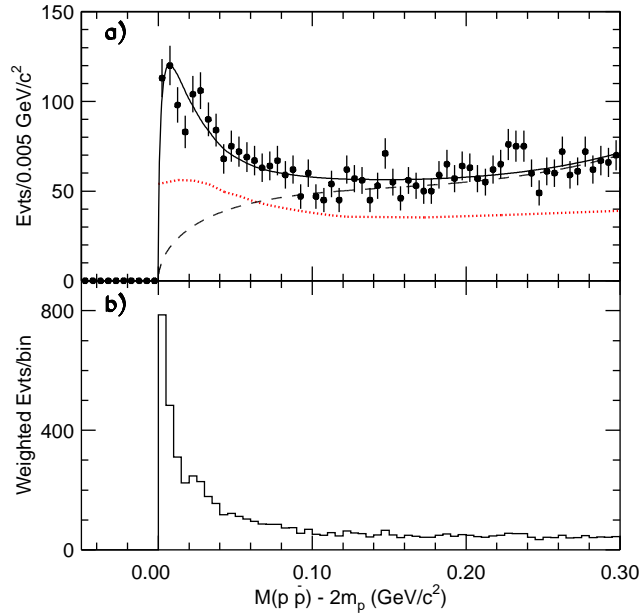}}
%\centerline{\psfig{file=fig1.eps,width=7cm}} \vspace*{8pt}
\caption{The threshold enhancement in the invariant mass spectrum
of $p \bar p$ pairs from Ref. \cite{bes}.\label{figure1} }
\end{figure}

BES Collaboration observed an enhancement near the threshold in
the invariant mass spectrum of $p \bar p$ pairs from $J/\psi \to
\gamma p \bar p$ radiative decays as shown in Fig. \ref{figure1}.
No similar structure is found in the $\pi^0 p\bar p$ channel
\cite{bes}. A naive S-wave Breit-Wigner fit yielded a central mass
below threshold
$M=1859^{+3}_{-10}(\mbox{stat})^{+5}_{-25}(\mbox{sys})$ MeV and
$\Gamma <30$ MeV \cite{bes}. There has been no report of the
similar signal in the $\pi\pi, \pi\pi\pi, K\bar K, K\bar K\pi$
channels, which suggests its quantum number be $J^{PC}=0^{-+},
I^G=0^+$ \cite{gao}, consistent with BES's S-wave fit.

\begin{figure}[th]
\scalebox{0.7}{\includegraphics{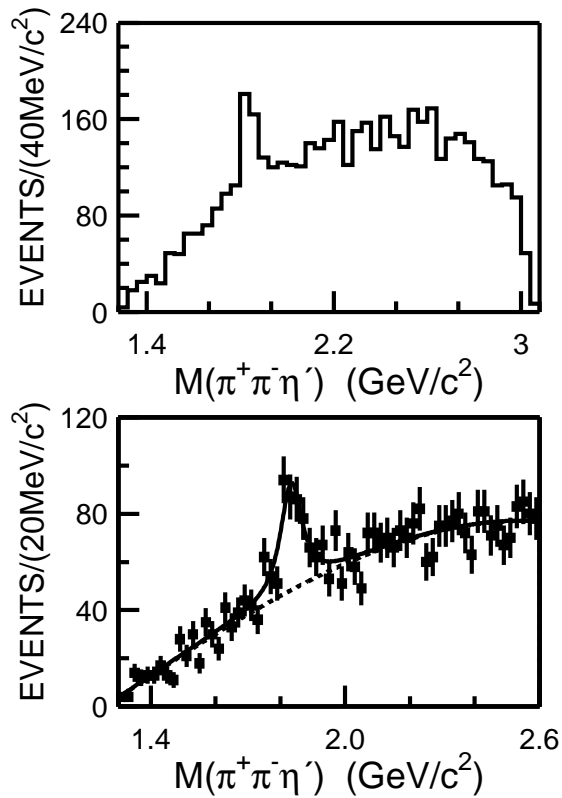}}
%\centerline{\psfig{file=fig2.eps,width=7cm}} \vspace*{8pt}
\caption{The invariant mass distribution of $\eta^\prime
\pi^+\pi^-$ from $J/\psi$ radiative decays in Ref.
\cite{bes1}.\label{figure2}}
\end{figure}

Later BES Collaboration reported a new signal X(1835) in the
$J/\psi \to\gamma X(1835) \to \gamma \eta^\prime \pi^+\pi^-$
channel \cite{bes1}. The invariant mass distribution of
$\eta^\prime \pi^+\pi^-$ is shown in Fig. \ref{figure2}. The
$\eta^\prime$ meson was detected in both $\eta\pi\pi$ and
$\gamma\rho$ channels. There are roughly $264\pm 54$ events. With
a statistical significance of $7.7\sigma$, the mass of X(1835) was
measured to be $(1833.7\pm 6.2\pm 2.7)$ MeV and its width to be
$(67.7\pm 20.3\pm 7.7)$ MeV \cite{bes1}. At the same time, the
$p\bar p$ enhancement was refit using the $p\bar p$ final state
interaction in the isoscalar channel obtained in Ref. \cite{fsi}
as input. Now the resulting mass was around 1830 MeV with a width
$\Gamma =(0\pm 93)$ MeV.

The mass and width of either the $p\bar p$ threshold enhancement
or $X(1835)$ do not match any known particle around this mass
range \cite{pdg}. There have been many speculations of the
underlying structure of the $p\bar p$ threshold enhancement and
X(1835) in literature. Proposed theoretical schemes include the
t-channel pion exchange, some kind of threshold kinematical
effects, a new resonance below threshold, even a $p\bar p$ bound
state etc
\cite{gao1,rosner,q1,q2,q3,q4,q5,q6,q7,q8,q9,q10,q11,q12,q13,q14,q15}.

The pseudoscalar glueball possibility was discussed in Refs.
\cite{koch,hxg,lba}. One potential obstacle of this assignment is
its low mass. Lattice QCD simulation predicts the pure scalar
glueball around $1.5\sim 1.7$ GeV \cite{lattice}. In the scalar
sector there are several possible experimental candidates around
$1.3\sim 1.7$ GeV \cite{pdg}. Pure pseudoscalar glueballs are
predicted to lie around 2.6 GeV \cite{lattice}. It seems a
powerful mixing mechanism is required to pull its mass from 2.6
GeV down to 1.835 GeV. One may also wonder why the dominant decay
mode of the pseudoscalar glueball is $p\bar p$.

Another very interesting possibility is the $p\bar p$ bound state
or baryonium if both the threshold enhancement and X(1835)
originate from a single genuine resonance. The study of nucleon
and anti-nucleon bound states dated back to Fermi and Yang
\cite{fermi}. An extensive theoretical and experimental review can
be found in Refs. \cite{history,pr}. With the help of G-parity,
the long and medium range part of the nucleon anti-nucleon
interaction $V_{N\bar N}$ may be related to that of $V_{N N}$. The
short-range part of $V_{N N}$ was well determined using the
properties of the light nuclei like deuteron and the enormous
nucleon-nucleon scattering data as inputs. In contrast, the
nucleon anti-nucleon scattering data is scarce. The short-range
part of $V_{N\bar N}$ remains essentially unknown. Especially the
annihilation contribution is very difficult to be taken into
account. Because of very poor knowledge of the short range part of
$V_{N\bar N}$, some deeply-bound $N\bar N$ states are always
predicted using the phenomenological $N\bar N$ potential.
Experimentally none of them was found. A reliable calculation of
the spectrum of $N\bar N$ bound states is still too demanding at
present.

Naively a $p\bar p$ baryonium may couple strongly with $p\bar p$
and have some good chance of decaying into $p\bar p$ through the
upper tail of the Breit-Wigner distribution if it's broad. A
$p\bar p$ bound state contains three up and down quark anti-quark
pairs. It could fall apart into three pairs of non-strange mesons
through color recombination. Since there is no valence strange
quark within the proton, the three-body strange decay channels
such as $K\bar K \pi$ are suppressed by OZI rule. The above two
simple observations seem consistent with BES's measurements.
However, one would simply expect the same signal in the
$\eta\pi\pi$ channel if the above picture is correct \cite{gao1},
unless there exists a special selection rule as found in Ref.
\cite{q6}.

However, there does not exist strong experimental evidence that
the $p\bar p$ threshold enhancement and X(1835) have the same
underlying structure. Very probably they are two different states
even if the enhancement arises from a sub-threshold resonance. For
example, the mass, total decay width, production rate and decay
pattern of X(1835) are consistent with its assignment as $\eta'$'s
second radial excitation \cite{tao}. Its decay mode $X(1835)\to
\eta'\pi^+\pi^-$ could occur through the emission of a pair of
S-wave pions, which is quite general for the double-pion decays of
ordinary radial excitations.

The clarification of the nature of both the $p\bar p$ enhancement
and X(1835) calls for more high-statistics data. Luckily BESIII
will start taking data this year. Investigation along this channel
will provide vital inputs for the poorly known nucleon and
anti-nucleon interaction, which is a fundamental issue in nuclear
and particle physics.

%%%%%%%%%%%%%%%%%%%%%%%%
\subsection{X(1576)}

\begin{figure}[th]
\scalebox{0.5}{\includegraphics{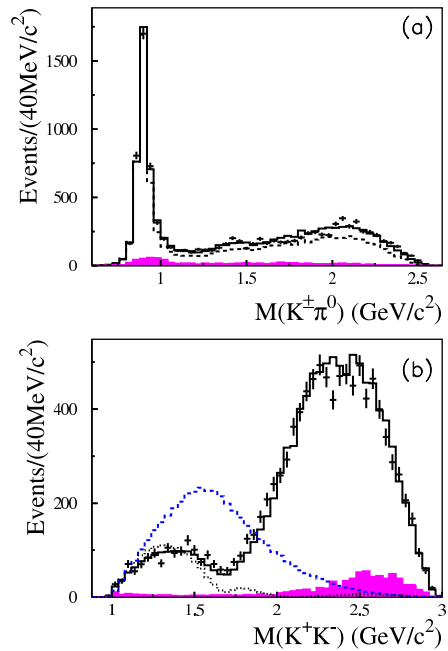}}
%\centerline{\psfig{file=fig3.eps,width=7cm}} \vspace*{8pt}
\caption{The $K^{+}K^{-}$ invariant mass spectrum in the
$J/\psi\to \pi^{0}K^{+}K^{-}$ channel from Ref.
\cite{1576-BES}.\label{figure3}}
\end{figure}

Last year BES Collaboration reported an extremely broad signal
$X(1576)$ in the $K^{+}K^{-}$ invariant mass spectrum in the
$J/\psi\to \pi^{0}K^{+}K^{-}$ channel \cite{1576-BES}, which is
shown in Fig. \ref{figure3}. Its quantum number and mass are
$J^{PC}I^G=1^{--}1^+$ and
$m=(1576^{+49}_{-55}(\mathrm{stat})^{+98}_{-91}(\mathrm{syst}))
-i(409^{+11}_{-12}(\mathrm{stat})^{+32}_{-67}(\mathrm{syst}))$ MeV
respectively. The branching ratio is $B[J/\psi\to
X(1576)\pi^{0}]\cdot B[X(1576)\to K^{+}K^{-}]=(8.5\pm
0.6^{+2.7}_{-3.6})\times 10^{-4}$.

Since there exist two broad resonances $\rho(1450)$ and
$\rho(1700)$ with exactly the same quantum numbers around 1.6 GeV,
one may wonder whether this signal arises from the interference of
these two nearby resonances. In fact, the interference effect
could produce an enhancement around $1540$ MeV in the $K^+K^-$
spectrum with typical interference phases. However, the branching
ratio $B[J/\psi\to \pi^{0}\rho(1450,1700)]\cdot
B[\rho(1450,1700)\to K^{+}K^{-}]$ from the final state interaction
effect is far less than the experimental data \cite{lx1}.

The large width around 800 MeV motivated theoretical speculations
that it could be a $K^{*}(892)-\kappa$ molecular state
\cite{Guo-1576}, tetraquark \cite{Lipkin-1576,1576-QSR},
diquark-antidiquark bound state \cite{Ding-1576,zhang}. If X(1576)
is a genuine resonance with $I=1$, its two charged partners await
to be discovered experimentally. One immediate challenge of both
the tetraquark and diquark-anti-diquark interpretations is where
to find the accompanying $SU_F(3)$ members within the same
multiplet.

As pointed out in Ref. \cite{Lipkin-1576}, the simple "fall-apart"
decay mechanism strongly suppresses the decay mode $X(1576)\to
\pi\pi$ and favors the $\pi\phi$ mode because of the presence of
the intrinsic $s\bar s$ pair if X(1576) is a tetraquark state with
the quark content $n\bar n s\bar s$ where $n$ denotes the up and
down quark. On the other hand, the $\pi\pi$ should be a favorable
mode while the $\pi\phi$ mode is suppressed by the OZI rule if
X(1576) is a conventional $q\bar q$ state with
$J^{PC}I^G=1^{--}1^+$. BES Collaboration did not report an
enhancement in the $\pi\phi$ spectrum in their previous study of
the $\pi\pi\phi$ channel \cite{bes3}.

%%%%%%%%%%%%%%%%%%%%%%%%%%%%%%%%%%
\subsection{$f_0(1810)$}

The low-lying scalar mesons have always been a big challenge to
hadron physicists. Up to now the underlying structure of the
$\sigma$, $\kappa$ and $f_0(980)$ mesons remain mysterious. Above
1 GeV there also exists an overpopulation of the scalar spectrum.
I.e., in the non-strange sector, there are $f_(1370)$,
$f_0(1500)$, $f_0(1710)$, $f_0(1790)$ \cite{pdg}.

\begin{figure}[th]
\scalebox{0.5}{\includegraphics{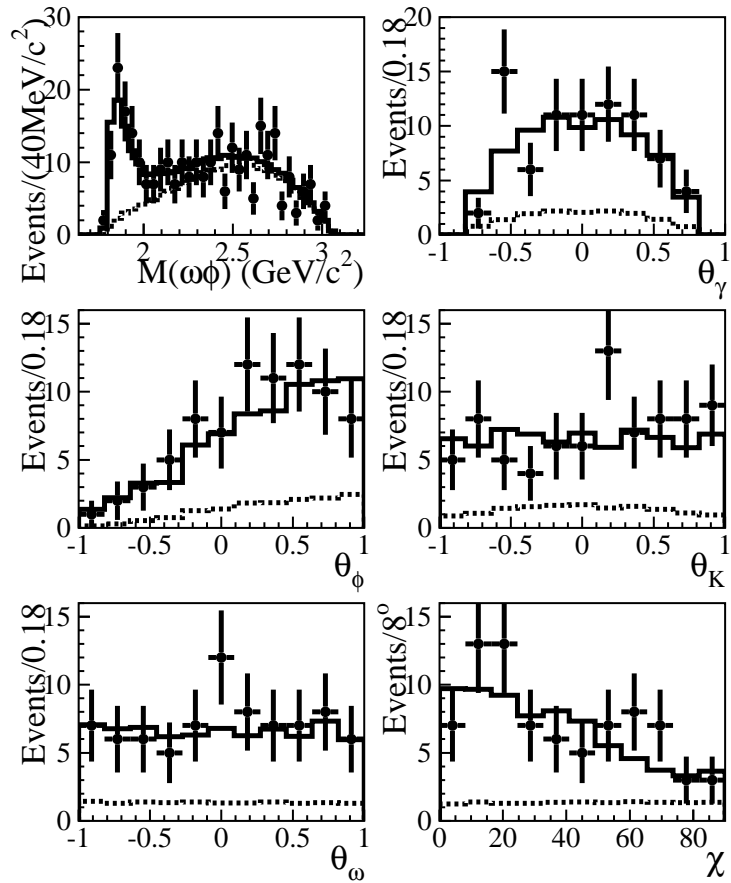}}
%\centerline{\psfig{file=fig4.eps,width=7cm}} \vspace*{8pt}
\caption{The $\omega \phi$ invariant mass spectrum from Ref.
\cite{bes4}.\label{figure4}}
\end{figure}

Recently a scalar signal near threshold $f_0(1810)$ or X(1810) is
observed in the $\omega \phi$ invariant mass spectrum (Fig.
\ref{figure4}) from the doubly OZI suppressed decays of $J/\psi
\to \gamma \omega \phi$ \cite{bes4}, which makes the "scalar
puzzle" more interesting.  A partial wave analysis shows that this
enhancement favors $J^P = 0^+$, and its mass and width are $M =
1812^{+19}_{-26} (stat) \pm 18 (syst) MeV/c^2$ and $\Gamma =
105\pm 20 (stat) \pm 28 (syst) MeV/c^2$. There is no report of the
same signal in the $\omega\omega$ channel. Experimentally it will
be very desirable to establish $f_0(1810)$ as an independent state
from the broad $f_0(1710)$ and $f_0(1790)$. It's hard to imagine
there exist five independent conventional $q\bar q$ scalars in the
interval $[1.3, 1.9]$ GeV, where one would expect a $s\bar s$
scalar accompanying $f_0(1370)$ if the latter one is a P-wave
$n\bar n$ state, and a third scalar which may be the radial
excitation of $f_0(1370)$. The overpopulation of the scalar
spectrum strongly indicates that there {\sl should} be some
non-$q\bar q$ components in some of these states.

Some authors propose X(1810) could be a scalar tetraquark state
\cite{tetra1,tetra2}. According to Ref. \cite{tetra1},
$X\to\omega\phi, K^* K^*$ are the two dominant decay channels
while $X\to KK, \eta\eta, \eta\eta'$ are suppressed. But a recent
analysis does not favor such an explanation \cite{tetra3}.

Chao proposed X(1810) as a scalar hybrid meson with the quark
content $n\bar n G_{TM}$ \cite{hybrid}. $G_{TM}$ is the transverse
gluon field in the language of the bag model, which couples to
$s\bar s$ more strongly than to $n\bar n$ \cite{hybrid1}. This
scheme provides a natural explanation of $\omega\phi$ as the
dominant decay mode. At the same time, X(1810) decays into
$\omega\omega, K^\ast \bar K^\ast$ with a smaller branching ratio.
Moreover, one inevitable prediction of this scheme is the
existence of the scalar hybrid nonet. Especially the other
isoscalar member $s\bar s G_{TM}$ decays mainly into $\phi\phi$ if
kinematically allowed, which can be tested by future BESIII data.

The authors of Ref. \cite{glue1} calculated the decay width of the
scalar glueball to be around 100 MeV, consistent with the width of
X(1810). They used the $K\bar K$ rescattering mechanism from the
open flavored decay channel to explain the flavor asymmetric
$\omega\phi$ decay mode of X(1810) and concluded that X(1810) is a
solid glueball candidate which can further be tested in the $K\bar
K$ channel.

Instead of the pure glueball interpretation, Bugg argued that the
$\omega\phi$ threshold peak X(1810) arises from a glueball
component in $f_0(1790)$ where the glueball distributes among
$f_(1370)$, $f_0(1500)$, $f_0(1710)$ and $f_0(1790)$ with
$f_0(1790)$ having a component of $40\%$ in intensity. A more
complicated mixing scheme was proposed in Ref. \cite{mix1}. There
are two P-wave quark model states $n\bar n$, $s\bar s$, two hybrid
scalar mesons $n\bar n G$, $s\bar s G$ before mixing. These five
"bare" states have the same quantum numbers and mix with each
other to produce $f_(1370)$, $f_0(1500)$, $f_0(1710)$, $f_0(1790)$
and X(1810).

Besides the above exotic schemes, some traditional interpretations
were also explored. Ref. \cite{threshold} suggests X(1810) could
arise from the S-wave threshold effects. Zhao and Zou noted that
the contributions from the vector meson $K^\ast \bar K^\ast$
rescattering via scalar meson exchange can produce some
enhancement near the $\omega\phi$ threshold \cite{zhao-fsi}.
However the resulting partial width of $X(1810)\to \omega\phi$
from the rescattering mechanism is much less than the experimental
data with large uncertainty.

%%%%%%%%%%%%%%%%%%%%%%%%%%%%%%%%%%%%%%%%%%%%%%%%%%%%%%%%%%%%
\subsection{Recent candidates of the $1^{-+}$ exotic mesons}

Hybrid mesons are composed of a pair of $q\bar q$ and one explicit
gluon field $G$. In the large $N_c$ limit, the amplitude of
creating a hybrid meson from the vacuum has the same $N_c$ order
as that of creating a $q\bar q$ meson \cite{cohen}. If kinematics
and other conservation laws allow, the production cross section of
hybrid mesons is expected to be roughly the same as that of
ordinary mesons. At least it's not suppressed in the large $N_c$
limit. In the same limit, hybrid mesons and ordinary mesons mix
freely if they carry the same quantum numbers. Hence, the
identification of hybrid mesons is very difficult unless they have
exotic quantum numbers. That's why so many efforts have been
devoted to the search of the $1^{-+}$ hybrid meson, which was
predicted to be the lightest exotic hybrid meson in the range of
$[1.9-2.1]$ GeV in some theoretical models.

Flux tube model predicts hybrid mesons prefer decaying into a pair
of mesons with L=1 and L=0 \cite{flux1,flux2}. Heavy hybrid mesons
tend to decay into one P-wave heavy meson and one pseudoscalar
meson according to a light-cone QCD sum rule calculation
\cite{zhu-qcd}. A lattice QCD simulation suggests the string
breaking mechanism may play an important role for the decays of
the hybrid heavy quarkonium \cite{bali}. When the string between
the heavy quark and anti-quark breaks, new light mesons are
created. In other words, the preferred final states are one heavy
quarkonium plus light mesons. However, readers should be very
cautious of these so-called "selection rules", none of which has
been tested by experiments because none of the $1^{-+}$ hybrid
candidates has been established unambiguously.

\begin{figure}[th]
\scalebox{0.5}{\includegraphics[angle=270]{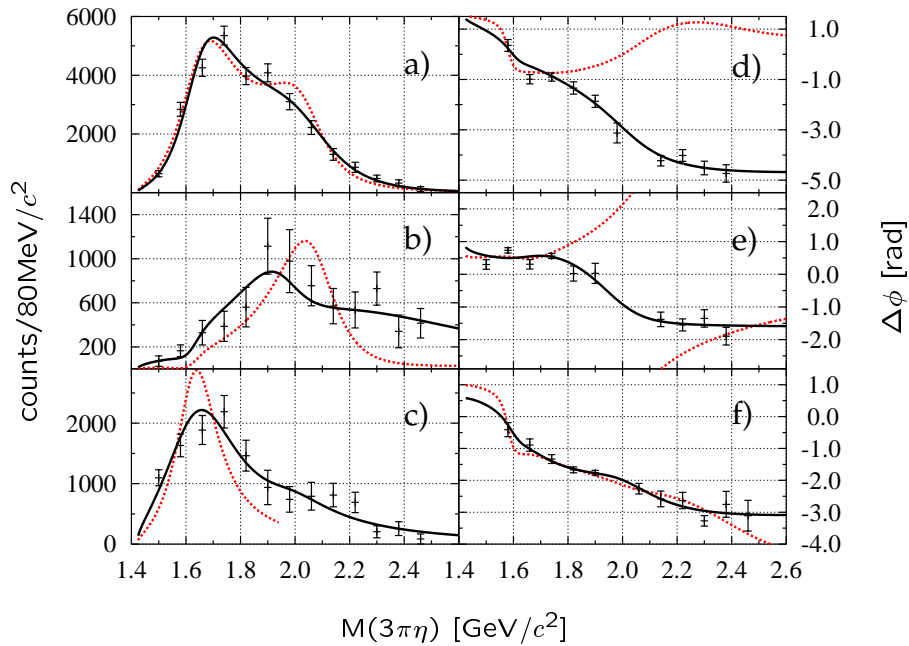}}
%\centerline{\psfig{file=fig5.eps,width=7cm}} \vspace*{8pt}
\caption{PWA results: $f_{1}(1285)\pi^{-}$ intensity distributions
 $1^{++}0^{+}f_{1}\pi^{-}P$, (b) $2^{-+}0^{+}f_{1}\pi^{-}D$, (c) $1^{-+}1^{+}f_{1}\pi^{-}S$
 and phase difference distributions (d) $\phi(1^{-+})-\phi(2^{-+})$,
 (e) $\phi(1^{-+})-\phi(1^{++})$, (f) $\phi(1^{++})-\phi(2^{-+})$ from Ref. \cite{f1pi}.
 \label{figure5}}
\end{figure}

Two $1^{-+}$ isovector states $\pi_1(1400)$ and $\pi_1(1600)$
appear in the PDG meson summary table \cite{pdg}. $\pi_1(1400)$
was observed in the $\eta\pi$ decay mode \cite{etapi1,etapi2}
while $\pi_1(1600)$ was observed in several channels $\rho\pi$
\cite{rhopi}, $\eta^\prime\pi$ \cite{etappi}, $f_1(1285)\pi$
\cite{f1pi}, $b_1(1235)\pi$ \cite{b1pi}. A third $1^{-+}$ state
$\pi_1(2000)$ was reported to have a width of 333 MeV in the
$f_1(1285)\pi$ (Fig. \ref{figure5}) \cite{f1pi} and $b_1(1235)\pi$
\cite{b1pi} channels.

\begin{figure}[th]
\scalebox{0.5}{\includegraphics{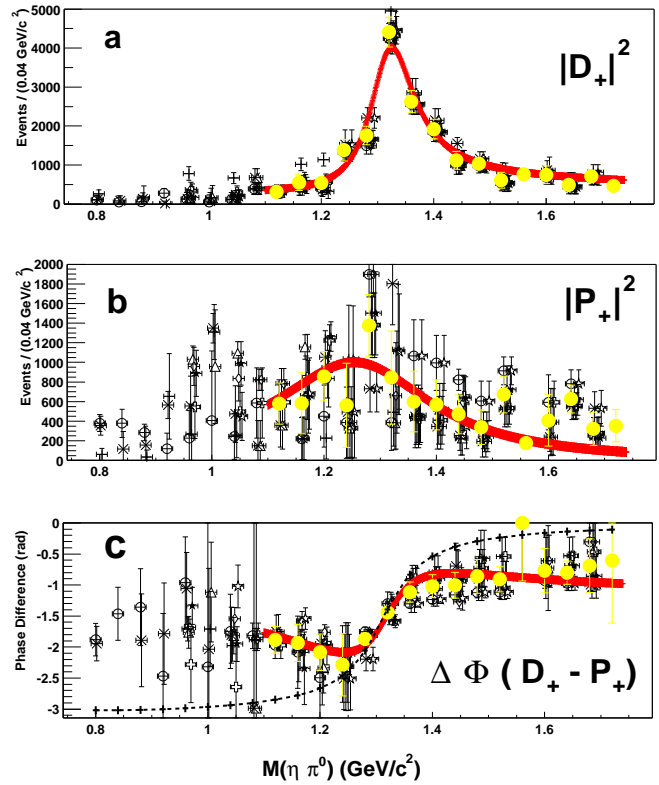}}
%\centerline{\psfig{file=fig7.eps,width=7cm}} \vspace*{8pt}
\caption{(a) The $D_+$ wave intensity; (b) the $P_+$ (or $1^{-+}$)
wave intensity; and (c) the relative phase between the $P_+$ and
$D_+$ waves from Ref. \cite{confirm}.\label{figure7}}
\end{figure}

Although there were concerns of the experimental analysis of the
$\pi_1(1400)$ signal \cite{challenge1}, the first state was
confirmed with a mass $1257 \pm 20 \pm 25$ MeV and width $354 \pm
64 \pm 60$ in the $\eta\pi^0$ mode (Fig. \ref{figure7})
\cite{confirm}.

A recent partial wave analysis of a higher statistics sample of
the $(3\pi)^-$ system in two charged modes found no evidence of an
exotic meson based on 3.0M $\pi^- \pi^0 \pi^0$ events and 2.6M
$\pi^- \pi^-\pi^+$ events \cite{challenge2}, which is in contrast
with an earlier analysis of 250K $\pi^- \pi^- \pi^+$ events from
the same experiment which showed possible evidence for a
$J^{PC}=1^{-+}$ exotic meson with a mass of $\sim$1.6 GeV decaying
into $\rho \pi$ \cite{rhopi}.

It's very encouraging to note that the mass and decay patterns of
$\pi_1(2000)$ are compatible with available theoretical
predictions. However, it's important to note that the gluon inside
the hybrid meson can easily split into a pair of $q\bar q$.
Therefore tetraquarks can always have the same quantum numbers as
the hybrid mesons, including the exotic ones. Discovery of hadron
candidates with $J^{PC}=1^{-+}$ does not ensure it's an exotic
hybrid meson. One has to exclude the tetraquark possibility based
on its mass, decay width and decay pattern etc. This argument
holds for all of the three exotic candidates $\Pi_1(1400)$,
$\Pi_1(1600)$ and $\Pi_1(2000)$.

%%%%%%%%%%%%%%%%%%%%%%%%%%%%%
\subsection{Y(2175)}

\begin{figure}[th]
\scalebox{0.5}{\includegraphics{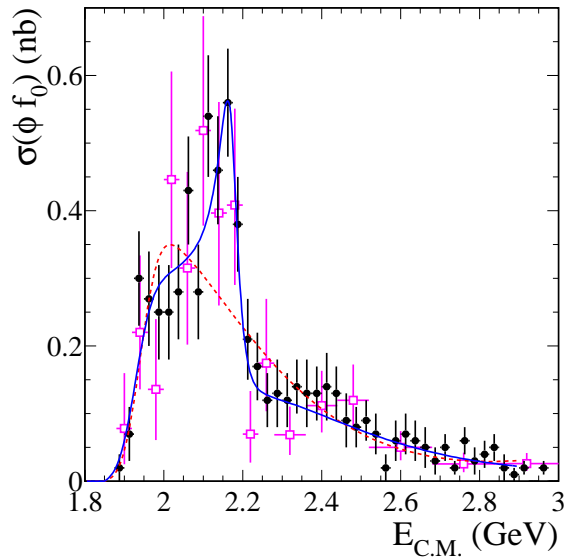}}
%\centerline{\psfig{file=fig8.eps,width=7cm}} \vspace*{8pt}
\caption{The threshold resonance Y(2175) from Ref.
\cite{babar10}.\label{figure8}}
\end{figure}

Babar collaboration observed a structure near threshold consistent
with a $1^{--}$ resonance with mass $m = 2.175 \pm 0.010\pm 0.015
$GeV and width $\Gamma = 58\pm 16\pm 20$ MeV (Fig. \ref{figure8})
in their measurement of the cross section for $e^+ e^- \to
\phi(1020) f_{0}(980)$ as a function of center-of-mass energy
using the initial-state-radiation technique \cite{babar10}. Since
this state was observed during Babar's search of the $Y(4260)\to
\phi\pi^+\pi^-$ decay mode, let's tentatively name it as Y(2175).
Y(4260) will be discussed in Section \ref{sec4}.

The possibility of Y(2175) being a tetraquark state with the quark
content $s\bar s s\bar s$ was studied in Ref. \cite{wzg10}.
However, Y(2175) is 175 MeV above the $\phi(1020) f_{0}(980)$
threshold. As a $s\bar s s\bar s$ tetraquark state, it will fall
apart into two mesons very easily and become very broad. The main
decay mode of Y(4260) should be $\phi(1020) f_{0}(980)$ if
$f_0(980)$ is a P-wave meson with a large (or dominant) $s\bar s$
component. C parity forbids the kinematically allowed modes
$\phi(1020) \phi(1020)$, $f_{0}(980)f_{0}(980)$. Other modes are
suppressed by the OZI rule.

It was proposed in Ref. \cite{ding10a} that Y(2175) could be a
strange hybrid meson. Its decay pattern and width were studied in
both the flux tube model and constituent gluon model. The total
width is generally larger than 100 MeV with some theoretical
uncertainty \cite{ding10a}. The lattice simulation suggests that
the strange $1^{-+}$ hybrid meson lies around 2175 MeV
\cite{lattice}.

Around its mass, there are two conventional $1^{--}$ $s\bar s$
states in the quark model, $2 ^3D_1$ and $3 ^3S_1$. According Ref.
\cite{barnes10}, the width of the $3^3S_1$ $s\bar s$ state is
about 380 MeV. The total width of the $2 ^3D_1$ state from both
$^3P_0$ and flux tube model is around $(150\sim 250)$ MeV
\cite{ding10b}. However, readers should keep in mind that the
predictions from these strong decay models sometimes deviate from
the experimental width by a factor of two or three. For
comparison, the widths of the $3^3S_1$ and $2 ^3D_1$ charmonium
are less than 110 MeV \cite{pdg}. Fortunately the characteristic
decay modes of Y(2175) as either a hybrid or $s\bar s$ state are
quite different, which may be used to distinguish the hybrid and
$s\bar s$ schemes. The possibility of Y(2175) arising from S-wave
threshold effects is not excluded.

%%%%%%%%%%%%%%
\subsection{$p\bar\Lambda$ threshold enhancement}

\begin{figure}[th]
\scalebox{0.5}{\includegraphics{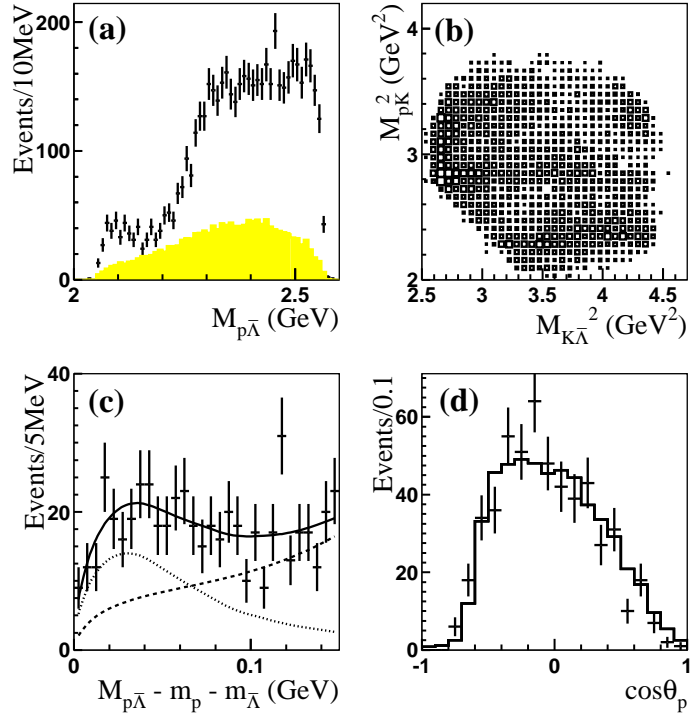}}
%\centerline{\psfig{file=fig9.eps,width=7cm}} \vspace*{8pt}
\caption{The $p\bar\Lambda$ threshold enhancement from Ref.
\cite{bes11}.\label{figure9}}
\end{figure}

BES collaboration observed an enhancement near the $m_p +
M_{\Lambda}$ mass threshold in the combined $p \bar{\Lambda}$ and
$\bar{p}\Lambda$ invariant mass spectrum (Fig. \ref{figure9}) from
$J/\psi \to p K^- \bar{\Lambda} + c.c. $ decays \cite{bes11}. If
fit with an S-wave Breit-Wigner resonance, its mass and width are
$m = 2075 \pm 12 (stat) \pm 5 (syst)$ MeV and $\Gamma = 90 \pm 35
(stat) \pm 9 (syst)$ MeV. There is also evidence for a similar
enhancement in $\psi' \to p K^- \bar{\Lambda} + c.c. $ decays.
Baryon anti-baryon low-mass enhancements were also observed in
charmless three-body baryonic B decays \cite{belle11}.

This threshold enhancement has been discussed in Refs.
\cite{he11,yuan11,ding11}. It was proposed as a baryon anti-baryon
bound state belonging to the SU(3) nonet \cite{yuan11}, similar to
Fermi and Yang's scheme \cite{fermi}. The problem with this
assignment is that two many baryon antibaryon bound states were
predicted, which were not confirmed by later experiments. The same
difficulty occurs with the interpretation of this enhancement as a
$\rm{q}^3\bar{\rm{q}}^3$ meson in the quark models with/without
interquark correlations \cite{ding11}.

%%%%%%%%%%%%%%%%%%%%%%%%%%%%%%%%%%%%
\section{Charmed mesons}\label{sec3}
%%%%%%%%%%%%%%%%%%%%%%%%%%%%%%%%%%%%

The heavy quark effective theory (HQET) provides a systematic
expansion in terms of $1/ m_Q$ for hadrons containing a single
heavy quark, where $m_Q$ is the heavy quark mass \cite{HQET}. The
angular monument of light components $j_{\ell}$ is a good quantum
number in the $m_Q\to\infty$ limit. Inside a heavy meson $q\bar
Q$, $j_{\ell} = L+S_q $ where $L$ and $S_q$ are the orbital and
spin angular momentum respectively. The heavy mesons can be
grouped into doublets with definite $j_{\ell}^P$. Except for
$L=0$, there are two doublets for $L\ge 1$. The mass splitting
within the same doublet scales like $1/m_Q$, hence vanishes in the
$m_Q\to\infty$ limit. The states with the same $J^P$, such as the
two $1^-$ and two $1^+$ states, can be distinguished in the
$m_Q\to\infty$ limit, which is one of the advantage of working in
HQET.

%%%%%%%%%%%%%%%%%%%%%%%%%%%%%%%%%%%%%%%%%%%%%%%%
\subsection{P-wave non-strange charmed mesons}

\begin{figure}[th]
\scalebox{0.5}{\includegraphics{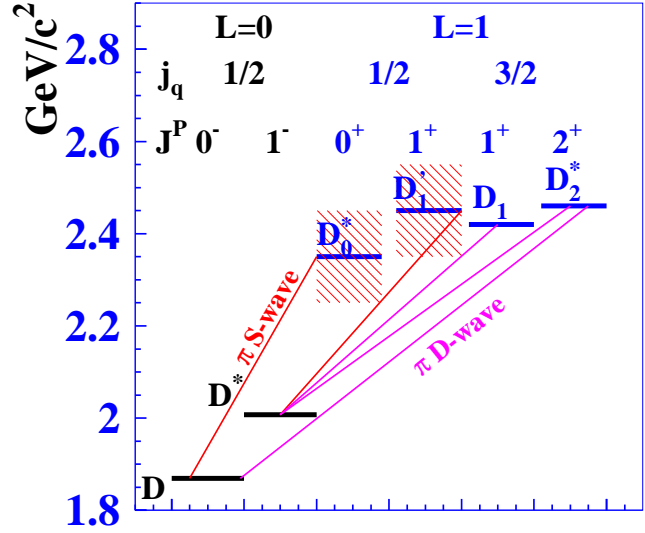}}
%\centerline{\psfig{file=fig100.eps,width=7cm}} \vspace*{8pt}
\caption{Energy levels of P-wave non-strange
mesons.\label{figure100}}
\end{figure}

The $L=0$, $j_{\ell} = \frac{1}{2}^-$ doublet $(0^-, 1^-)$
corresponds to the pseudoscalar and vector ground state mesons.
For $L=1$, $j_{\ell} = \frac{1}{2}^+$ or $j_{\ell} =
\frac{3}{2}^+$. In the heavy quark limit, the strong decays of
heavy mesons conserve parity, the total angular momentum and the
angular momentum of the light components. The $\frac{1}{2}^+$
doublet $(0^+,1^+)$ mainly decays into a ground state heavy meson
plus a pseudoscalar light meson through S-wave, if kinematically
allowed. The $\frac{3}{2}^+$ doublet $(1^+,2^+)$ decays through
D-wave. Hence the $(0^+,1^+)$ doublet has a broad width around
several hundred MeV while the $(1^+,2^+)$ doublet is quite narrow
with a width less than 50 MeV \cite{pdg}. The energy levels of the
ground state and P-wave charmed mesons are shown in Fig.
\ref{figure100}. For $L=2$ there are $(1^{-},2^{-})$ and
$(2^{-},3^{-})$ doublets with $j_{\ell}^P=\frac{3}{2}^-$and
$\frac{5}{2}^-$ respectively.

At present there are two puzzles in the system of P-wave charmed
mesons. The first puzzle is the near degeneracy of the non-strange
and strange $(0^+,1^+)$ doublet. For comparison, we collect the
masses of the $(0^-, 1^-)$ and $(1^+,2^+)$ doublets below. For the
non-strange D mesons, we have $m_D=1869$ MeV, $m_{D^\ast}=2010$
MeV, $m_{D_1}=2420$ MeV, $m_{D_2}=2460$ MeV. For their strange
counterparts, we have $m_{D_s}=1968$ MeV, $m_{D_s^\ast}=2112$ MeV,
$m_{D_{s1}}=2536$ MeV, $m_{D_{s2}}=2573$ MeV. The mass splitting
between the non-strange and strange $(0^-, 1^-)$ and $(1^+,2^+)$
doublets is roughly 100 MeV and 113 MeV respectively, completely
consistent with the naive expectation of $m_s-m_u\sim 110$ MeV.

\begin{figure}[th]
\scalebox{0.4}{\includegraphics{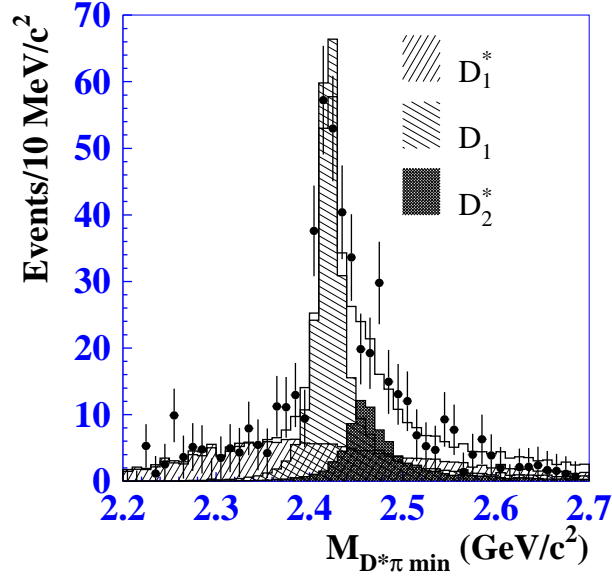}}
%\centerline{\psfig{file=fig10.eps,width=7cm}} \vspace*{8pt}
\caption{$D^\ast_{1}$ in the $D^\ast \pi$ spectrum from Ref.
\cite{focus1}.\label{figure10}}
\end{figure}

\begin{figure}[th]
\scalebox{0.4}{\includegraphics{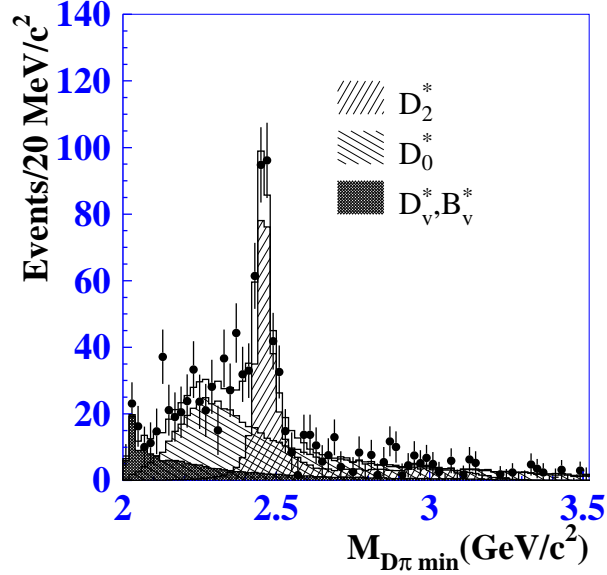}}
%\centerline{\psfig{file=fig15.eps,width=10cm}} \vspace*{8pt}
\caption{$D^\ast_{0}$ in the $D \pi$ spectrum from Ref.
\cite{focus1}.\label{figure12}}
\end{figure}

Now let's move on to the $(0^+,1^+)$ doublet, where the situation
seems very different according to the available experimental data
\cite{pdg}. For the strange doublet, we have
$m_{D^\ast_{s0}}=2317$ MeV, $m_{D^\ast_{s1}}=2459$ MeV, where we
use the star to indicate this $1^+$ state belongs to the
$(0^+,1^+)$ doublet. They are very narrow states with a width less
than 5 MeV. For the non-strange doublet there is one measurement
of the $1^+$ mass from Belle Collaboration $m_{D^\ast_{1}}=2427\pm
26\pm 25$ MeV with a large width $384^{+107}_{-75}\pm 74$ MeV
(Fig. \ref{figure10}) \cite{focus1}. For the $0^+$ state, there
were two measurements. FOCUS Collaborations reported
$m_{D^\ast_{0}}=2308 \pm 17 \pm 32$ MeV with a width $276\pm 21\pm
63 $ MeV \cite{belle1} while BELLE observed it at $2407\pm 21\pm
35$ MeV with a width $240\pm 55 \pm 59$ MeV (Fig. \ref{figure12})
\cite{focus1}. Because both $D^\ast_0$ and $D^\ast_1$ are broad
resonances, the experimental extraction of their mass and width is
difficult. Clearly the strange and non-strange $(0^+,1^+)$
doublets seem degenerate although with the large uncertainty,
which is reminiscent of the similar degeneracy among the light
strange and non-strange scalar mesons around 1.4 GeV such as
$a_0(1450)$ and $K_0^\ast (1430)$.

In fact the near-degeneracy of the strange and non-strange
$(0^+,1^+)$ doublets motivated Dmitrasinovic to propose the
tetraquark scheme \cite{pol}. It's quite natural to explain their
degenerate masses if one puts $D_0^\ast$ and $D_{s0}^\ast$ in the
anti-symmetric flavor SU(3) multiplet $\bar 3_A$ with the flavor
wave functions $|D_0^\ast\rangle = {1\over 2} | c (s(\bar u\bar
s-\bar s\bar u)-d (\bar d\bar u-\bar u\bar d)) \rangle $,
$|D_{s0}^\ast\rangle = {1\over 2} | c (u(\bar u\bar s-\bar s\bar
u)-d (\bar d\bar s-\bar s\bar d)) \rangle $. Unfortunately, as
always is the case, the $\bar 3_A$ tetraquark multiplet is
accompanied by a symmetric $\bar 3_S$ and ${\bar 15}$. The
additional states have not been found by experiments yet.

%%%%%%%%%%%%%%%%%%%%%%%%%%%%%%%%%%%%%%%%%%%%%%%%%%%
\subsection{$D_{sj}(2317)$ and $D_{sj}(2460)$}

\subsubsection{Interpretations of $D_{sj}(2317)$ and $D_{sj}(2460)$}

\begin{figure}[th]
\scalebox{0.5}{\includegraphics{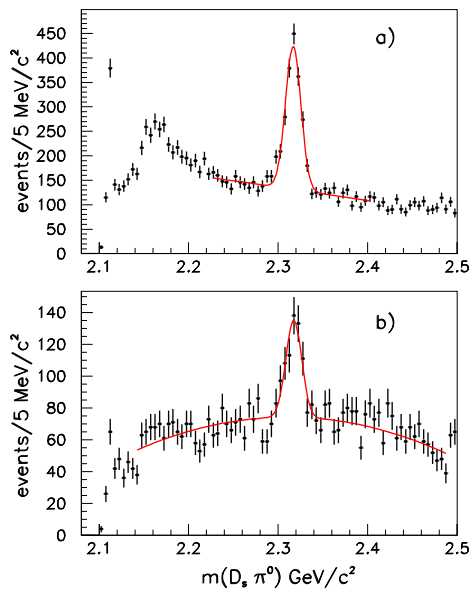}}
%\centerline{\psfig{file=fig13.eps,width=7cm}} \vspace*{8pt}
\caption{The narrow resonance $D_{sj}(2317)$ in the $D_s\pi^0$
channel in Ref. \cite{babar1}.\label{figure13}}
\end{figure}

\begin{figure}[th]
\scalebox{0.5}{\includegraphics{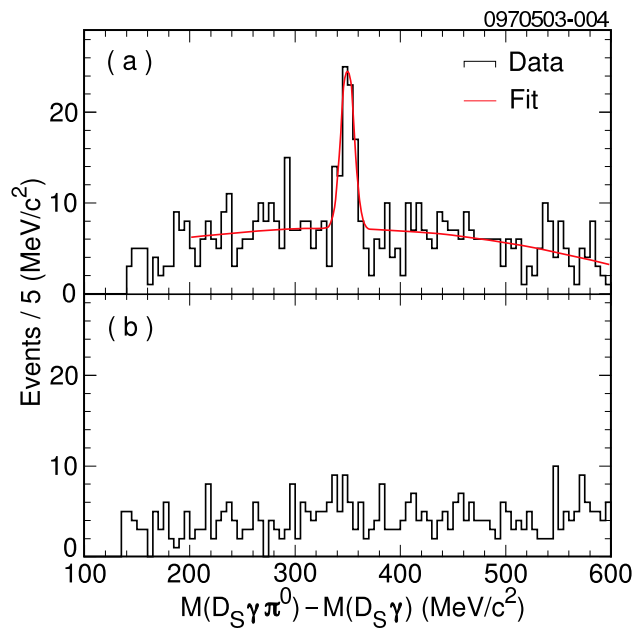}}
%\centerline{\psfig{file=fig14.eps,width=7cm}} \vspace*{8pt}
\caption{$D_{sj}(2460)$ in the $D^*_s\pi^0$ channel from Ref.
\cite{cleo}.\label{figure14}}
\end{figure}

In 2003 BaBar Collaboration discovered a positive-parity scalar
charm strange meson $D_{sJ}(2317)$ with a very narrow width in the
$D_s\pi^0$ channel (Fig. \ref{figure13}) \cite{babar1}, which was
confirmed by CLEO later \cite{cleo}. In the same experiment CLEO
observed the $1^+$ partner state at $2460$ MeV in the $D^*_s\pi^0$
channel (Fig. \ref{figure14}) \cite{cleo}. Up to now there have
been lots of experimental investigations of these two narrow
resonances
\cite{belle1b,belle2,belle3,belle4,focus,babar2,babar3,babar4,babar5}.

These two states lie below $DK$ and $D^\ast K$ threshold
respectively. The potentially dominant s-wave decay modes
$D_{sJ}(2317) \to D_sK$ etc are kinematically forbidden. Thus the
radiative decays and isospin-violating strong decays become
dominant decay modes. Therefore they are extremely narrow.

The discovery of these two states has triggered heated discussion
on their nature in literature. The key point is to understand
their low masses. Although it's tempting to classify these two
states as the $(0^+, 1^+)$ P-wave $c\bar{s}$ doublet, their masses
are 100 MeV lower than the expected values in quark models
\cite{qm}. The model using the heavy quark mass expansion of the
relativistic Bethe-Salpeter equation predicted a lower value
$2.369$ GeV for $D_{sJ}(2317)$ \cite{jin}, which is still 50 MeV
higher than the experimental data.

The mass of the $D_s(0^+)$ state from the lattice QCD calculation
is significantly larger than the experimentally observed mass of
$D_{sJ}(2317)$ \cite{bali2,dougall,soni}. It is pointed out
\cite{bali} that $D_{sJ}(2317)$ might receive a large component of
$DK$. Such a large $DK$ component makes lattice simulation very
difficult.

Barnes, Close and Lipkin argued that $D_{sJ}(2317)$ is a  DK
molecule at 2.32 GeV while the higher-mass scalar $c\bar s$ state
lies at 2.48 GeV \cite{barnes}. This $c\bar s$ state at 2.48 GeV
couples to the D K channel strongly, which would encourage
formation of an I=0 DK molecule. In order to explain its narrow
width and the isospin violating decay mode $D_s \pi$,
$D_{sJ}(2317)$ is proposed to be a dominantly I=0 DK state with
some I=1 admixture.

Van Beveren and Rupp \cite{rupp} argued from the experience with
$a_0/f_0 (980)$ that the low mass of $D_{sJ}(2317)$ could arise
from the mixing between the $0^+$ $\bar{c}s$ state and the $DK$
continuum. In this way the lowest $0^+$ state is pushed much lower
than that expected from quark models. Within their unitarized
meson model, $D^*_{sJ}(2317)$ is described as a {\sl quasi-bound}
scalar $c\bar{s}$ state owing its existence to the strong coupling
to the nearby S-wave DK threshold. The standard $c\bar{s}$ charm
strange scalar meson $D_{s0}$ is predicted around 2.79 GeV with a
width of 200 MeV, which is the analogue of the scalar nonet
$f_0(1370)$, $f_0(1500)$, $K_0^*(1430)$, and $a_0(1450)$.

If the chiral symmetry is linearly realized in the form of
$SU(3)_L \times SU(3)_R$ in the heavy-light meson systems, the
ground state $(0^-, 1^-)$ heavy multiplet will be accompanied by
its $(0^+, 1^+)$ heavy doublet. Bardeen, Eichten and Hill
interpreted $D_{sJ}(2317)$ and $D_{sJ}(2460)$ as the $c \bar s$
$(0^+, 1^+)$ spin parity partners of the $(0^-, 1^-)$ doublet in
the framework of chiral symmetry \cite{bardeen}.

There were also suggestions that $D_{sJ}(2317)$ and $D_{sJ}(2460)$
could be members of a host of $c\bar{s} n \bar{n}$ tetraquark
states \cite{cheng,kt}. However, quark model calculations show
that the mass of the four quark state is much larger than the
$0^+$ $\bar{c}s$ state \cite{vijande,zzy}. Furthermore, the four
quark system has five spin states. As always, the two demanding
issues of the tetraquark interpretations remain: (1) where are the
traditional $c\bar s$ $(0^+, 1^+)$ states in the quark model? (2)
where are those additional broad member states within the same
multiplet?

Up to now, only two states have been found after the experimental
scan in the mass range [2.3, 2.85] GeV, which is consistent with
the $\bar c s$ interpretation \cite{col1,godfrey,dai1} and
challenges those theoretical schemes predicting additional $c\bar
s$ states.

\subsubsection{Radiative decays of $D_{sJ}(2317)$ and $D_{sJ}(2460)$}

Radiative decays of $D_{sJ}(2317)$ and $D_{sJ}(2460)$ mesons were
suggested to explore their underlying structure in Ref.
\cite{godfrey}. Their electromagnetic widths were calculated after
considering the possible mixing between the two $1^+$ states and
assuming they are conventional $c\bar s$ states \cite{godfrey}.
Under the same assumption, the radiative decay width of
$D_{sJ}(2317)$ was estimated with the help of heavy quark symmetry
and the vector meson dominance model \cite{col1}. Later the
radiative decays were studied in the framework of light-cone QCD
sum rules (LCQSR). The numerical results favor the interpretation
of $D_{sJ}(2317)$ and $D_{sJ}(2460)$ as ordinary $\bar c s$ mesons
\cite{col2}. We collect these results in Table \ref{tab-rad}.
Experimentally only $D_{sJ}(2460)\rightarrow D_s \gamma$ has been
observed by Belle \cite{belle1b,belle2} and Babar
\cite{babar2,babar5} Collaborations.

\begin{table}[h]
\caption{Radiative decay widths of $D_{sJ}(2317)$ and
$D_{sJ}(2460)$ from various theoretical approaches in unit of
keV.} \label{tab-rad}
\begin{center}
\begin{tabular}{|c|c|c|c|}\hline
 References & \cite{godfrey}& \cite{col1}&
\cite{col2}\\ \hline
 $\Gamma (D_{sJ}(2317)\rightarrow D^{*}_s+\gamma)$& 1.9&1&4-6\\
 $\Gamma(D_{sJ}(2460)\rightarrow D_s \gamma)$&6.2&-&19-29\\
 $\Gamma(D_{sJ}(2460)\rightarrow D^{*}_s+\gamma)$&5.5&-&0.6-1.1\\
 $ \Gamma(D_{sJ}(2460)\rightarrow D_{sJ}(2317)+\gamma)$&-&-&0.5-1.8\\ \hline
\end{tabular}
\end{center}
\end{table}

\subsubsection{Strong decays of $D_{sJ}(2317)$ and $D_{sJ}(2460)$}

The decay channels $D_{sj}(2317)\rightarrow D\;K$ and
$D_{sj}(2460)\rightarrow D^*\;K$ are forbidden by kinematics.
Therefore, their possible strong decay modes are one-pion and
two-pion decays. The two-pion decay occurs via a virtual meson
such as $f_0(980)$. The one-pion decay mode breaks the isospin
symmetry and happens through $\eta$-$\pi^0$ mixing \cite{wise}:
$D_{sj}(2317)\rightarrow D_s\eta\rightarrow D_s\pi^0$,
$D_{sj}(2460)\rightarrow D^*_s\eta\rightarrow D^*_s\pi^0$. The
$\eta-\pi^0$ mixing is described by the isospin violating piece in
the chiral lagrangian
\begin{equation}
{\cal{L}}_m=\frac{m_{\pi}^2 f^2}{4(m_u+m_d)}\mbox{Tr}(\xi m_q
\xi+\xi^{\dagger} m_q \xi^{\dagger})~,
\end{equation}
where $\xi= \exp (i\tilde{\pi}/f_\pi)$,  $\tilde{\pi}$ the light
meson octet and $m_q$ is the light quark mass matrix. Such a
mixing is suppressed by the factor
$\frac{m_d-m_u}{m_s-\frac{m_u+m_d}{2}}$. Numerically the isospin
violating effect is ${\cal O}(10^{-2})$ in the amplitude. While
the isospin conserving strong decay width is ${\cal O}(10^{2})$
MeV, one would naturally expect the one-pion decay width of
$D_{sj}(2317)$ and $D_{sj}(2460)$ to be around several tens keV.

With the identification of ($D_{sJ}(2317), D_{sJ}(2460)$) as the
($0^+$, $1^+$) doublet in the heavy quark effective field theory,
the light cone QCD sum rule is derived for the coupling of eta
meson with $D_{sJ}(2317) D_s $ and $D_{sJ}(2460) D_s^{*} $.
Through $\eta-\pi^0$ mixing, their pionic decay widths are derived
\cite{wei1}. These two widths are similar in magnitude, as
expected from heavy quark symmetry. Combining the radiative decay
widths derived by Colangelo, Fazio and Ozpineci in the same
framework \cite{col2}, the authors conclude that the decay
patterns of $D_{sJ}(2317, 2460)$ strongly support their
interpretation as ordinary $c \bar s$ mesons.

In Ref. \cite{lu} pionic decay widths of $D_{sj}(2317)$ and
$D_{sj}(2460)$ were estimated using the $^3P_0$ model. Their
one-pion decays occur through $\eta$-$\pi^0$ mixing. The mixing
between $^3P_1$ and $^1P_1$ states enhances the single pion decay
width of $D_{sj}(2460)$. The double pion decays of $D_{sj}(2460)$
are allowed by isospin symmetry. But they are suppressed by
three-body phase space. Under a rather crude assumption that such
decays occur with the help of a virtual $f_0(980)$ meson, the
double pion decay width was estimated to be around $0.9\sim 1.1$
keV for $D_{sj}(2460) \to D_s + 2\pi$ mode and $(0.3\sim 0.7)$ keV
for $D_{sj}(2460) \to D^*_s + 2\pi$ mode, depending on the total
pionic width of $f_0(980)$. The double pion decay width of
$D_{sj}(2460)$ is numerically much smaller than its single pion
width because of the cancellation from the mixing of $^3P_1$ and
$^1P_1$ states. Putting the single and double pion decay modes
together, the strong decay width of $D_{sj}(2317, 2460)$ is less
than 50 keV. For comparison, the single pionic decay widths from
various theoretcal schemes were collected in Table
\ref{tab-strong}.

\begin{table}[h]
\caption{\baselineskip 15pt Single-pion decay widths (in keV) of
$D_{sJ}(2317)$ and $D_{sJ}(2460)$ mesons from various theoretical
approaches.} \label{tab-strong}
\begin{center}
\begin{tabular}
{c c c c c c c c c c} \hline References &  \cite{lu}&  \cite{wei1}
& \cite{col1} &
\cite{bardeen} & \cite{godfrey} &\cite{faya}&\cite{cheng}&\cite{ishida}&\\
\hline
$D^*_{sJ}(2317)\rightarrow D_{s}\pi^0$& 32 &  34-44   & $7\pm 1  $& 21.5  &$\sim 10 $& 16 & 10-100&$150\pm 70$\\
$D_{sJ}(2460)\rightarrow D_{s}^{*}\pi^0$& 35 & 35-51  & $7\pm 1  $& 21.5  &$\sim 10 $& 32 &       &$150\pm 70$\\
\hline
\end{tabular}
\end{center}
\end{table}

The experimental ratio of radiative and strong decay widths of
$D_{sJ}$ mesons is presented in Table \ref{tab-ratio} together
with the central values of theoretical predictions from light-cone
QCD sum rules. The theoretical ratio is consistent with Belle and
Babar's most recent data, which strongly indicates $D_{sJ}(2317)$
and $D_{sJ}(2460)$ are conventional $c\bar s$ mesons.

\begin{table}[h]
\caption{Comparison between experimental ratio of $D_{sJ}(2317,
2460)$ radiative and strong decay widths and theoretical
predictions from light-cone QCD sum rule approach.}
\label{tab-ratio}
\begin{center}
\begin{tabular}{cccc|c}
\hline & Belle & Babar   & CLEO  & LCQSR\\ \hline $\frac{\Gamma
\left( D^*_{sJ}(2317) \rightarrow D_{s}^{\ast }\gamma \right) }{
\Gamma \left( D^*_{sJ}(2317)\rightarrow D_{s}\pi ^{0}\right) }$ &
$<0.18$ \cite{belle2}& & $<0.059$& 0.13 \\
\hline $\frac{\Gamma \left(D_{sJ}(2460) \rightarrow D_{s}\gamma
\right) }{ \Gamma \left( D_{sJ}(2460)\rightarrow D_{s}^{\ast }\pi
^{0}\right) }$ & $0.55\pm0.13$  & $0.375\pm0.054$
&$<0.49$ & 0.56\\
& $\pm0.08$ \cite{belle2} & $\pm0.057$ \cite{babar5} & &\\
\hline $\frac{\Gamma \left( D_{sJ}(2460)\rightarrow D_{s}^{\ast
}\gamma \right) }{\Gamma \left( D_{sJ}(2460)\rightarrow
D_{s}^{\ast }\pi ^{0}\right) }$
& $<0.31$ \cite{belle2}& &$<0.16$ & 0.02 \\
\hline $\frac{\Gamma \left( D_{sJ}(2460)\rightarrow
D^*_{sJ}(2317)\gamma \right) }{\Gamma \left(
D_{sJ}(2460)\rightarrow
D_{s}^*\pi^0 \right) }$  &  & $ < 0.23$   \cite{babar4}& $ < 0.58$ & 0.015 \\
\hline
\end{tabular}
\end{center}
\end{table}

Combining the radiative decay width, the total width of
$D_{sj}(2317, 2460)$ is less than 100 keV. Both resonances are
extremely narrow if they are $c\bar s$ states. A precise
measurement of their total widths may help distinguish theoretical
models of their quark content. In the future, B decays into
$D_{sJ}$ mesons may also play an important role in exploring these
charming states.

\subsubsection{The low mass puzzle of $D_{sj}(2317)$ and $D_{sj}(2460)$}

The masses of $D_{sj}(2317)$ and $D_{sj}(2460)$ have been treated
with QCD sum rules in heavy quark effective theory in \cite{dai1}.
The result for the the $D_s(0^+)$ mass is consistent with
experimental data within the large theoretical uncertainty.
However, the central value is still $90 $ MeV larger than the
data. Even larger result for the $D_s(0^+)$ mass was obtained in
the earlier work with the sum rule in full QCD \cite{colangelo91}.
It has been pointed out in \cite{dai1} that in the formalism of
QCD sum rules the physics of mixing with $DK$ continuum resides in
the contribution of $DK$ continuum in the sum rule and including
this contribution should render the mass of $D_s(0^+)$ lower.

There have been two investigations on this problem using sum rules
in full QCD including the $O(\alpha_s)$ corrections. In Ref.
\cite{haya} the value of the charm quark pole mass $M_c=1.46 $ GeV
is used and $0^+ \bar{c}s$ is found to be $100-200$ MeV higher
than the experimental data. On the other hand, in Ref.
\cite{narison} $M_c\simeq 1.33 $GeV is used and the central value
of the results for the $0^+ \bar{c}s$ mass is in good agreement
with the data. As commented by the author, the uncertainty in the
value of $M_c$ is large.

The contribution of the two-particle continuum to the spectral
density can safely be neglected in many cases, as usually done in
the traditional QCD sum rule analysis. One typical example is the
rho meson sum rule, where the two pion continuum is of p-wave. Its
contribution to the spectral density is tiny and the rho pole
contribution dominates.

However, there may be an exception when the $0^+$ particle couples
strongly to the two particle continuum via s-wave. In such case,
there is no threshold suppression and the two-particle continuum
contribution may be more significant. The strong coupling of the
$0^+$ particle with the two particle state and the adjacency of
the $0^+$ mass to the DK continuum threshold result in large
coupling channel effect which corresponds to the configuration
mixing in the formalism of the quark model. In the problem under
consideration, $D_{sJ}(2317)$ is only 48 MeV below DK threshold
and the s wave coupling of $D_s(0^+)DK$ is found to be very large
\cite{colangelo95,zhuzhu}. Therefore, one may have to take into
account the $DK$ continuum contribution carefully. The importance
of the $D\pi$ continuum in the sum rule for the $0^+$ particle was
first emphasized in Ref. \cite{shifman}, based on duality
consideration in the case where $0^+$ mass is higher than the
threshold. A crude analysis of the $B\pi$ continuum contribution
was made, based on the soft pion theorem in the case where $0^+$
mass is higher than the threshold \cite{zhuzhu}. Such an mechanism
may also explain partly why the extracted mass of the $0^+ \bar c
s$ state from the quenched lattice QCD simulation is higher than
the experimental value where the DK continuum contribution was not
included.

%%%%%%%%%%%%%%%%%%%%%%%%%%%%%%%%%%%%%%%%%%%%%%%%%%%%%%%%%%%%%%%%
\subsection{Recent candidates of higher excited charmed mesons}

\begin{figure}[th]
\scalebox{0.7}{\includegraphics{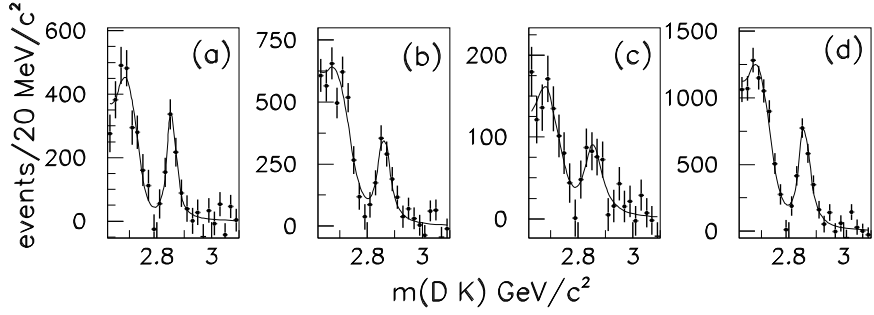}}
%\centerline{\psfig{file=fig15.eps,width=12cm}} \vspace*{8pt}
\caption{$D_{sJ}(2860)$ in the $DK$ invariant mass distribution
from Ref. \cite{babar-high}.\label{figure15}}
\end{figure}

\begin{figure}[th]
\scalebox{0.4}{\includegraphics{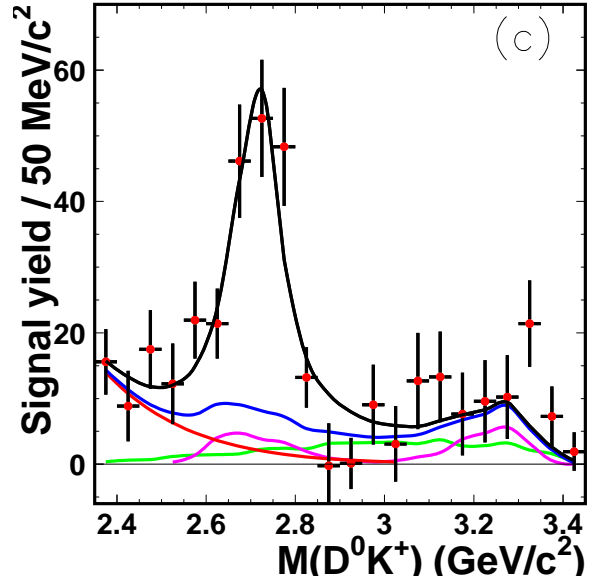}}
%\centerline{\psfig{file=fig16.eps,width=7cm}} \vspace*{8pt}
\caption{$D_{sJ}(2715)$ in the $D^{0}K^{+}$ spectrum from Ref.
\cite{belle-high}.\label{figure16}}
\end{figure}

The two d-wave $c\bar s$ doublets $(1^-, 2^-)$ and $(2^-, 3^-)$
have not been observed experimentally. Recently BaBar
Collaborations reported two new $D_{s}$ states, $D_{sJ}(2860)$
(Fig. \ref{figure15}) and $D_{sJ}(2690)$ in the $DK$ channel.
Their widths are $\Gamma=48\pm7\pm10 $ MeV and
$\Gamma=112\pm7\pm36 $ MeV respectively \cite{babar-high}. Thus
its $J^{P}=0^{+}, 1^{-}, 2^{+}, 3^{-}, \cdots$. At the same time,
Belle collaboration reported a broad $c\bar{s}$ state
$D_{sJ}(2715)$ (Fig. \ref{figure16}) with $J^{P}=1^{-}$ in
$B^{+}\to \bar{D}^{0}D^{0}K^{+}$ decay \cite{belle-high}. Its mass
is $2715\pm 11^{+11}_{-14}$ MeV and width $\Gamma=(115\pm
20^{+36}_{-32})$ MeV.

The $J^P$ of $D_{sJ}(2860)$ and $D_{sJ}(2690)$ can be $0^+,1^-,
2^+,3^-,\cdots$ since they decay into two pseudoscalar mesons.
$D_{sJ}(2860)$ was proposed as the first radial excitation of
$D_{sJ}(2317)$ based on a coupled channel model \cite{rupp3} or
the first radial excitation of the ground $1^-$ state $D_s^{*}$
using an improved potential model \cite{close3}. Colangelo et al
considered $D_{sJ}(2860)$ as the D wave $3^{-}$ state
\cite{colangelo3}. The mass of $D_{sJ}(2715)$ or $D_{sJ}(2690)$ is
consistent with the potential model prediction of  the $c\bar{s}$
radially excited $2^3S_1$ state \cite{isgur3,close3}. Based on
chiral symmetry consideration, a D wave $1^-$ state with mass
$M=2720 $ MeV is also predicted if the $D_{sJ}(2536)$ is taken as
the P wave $1^+$ state \cite{nowak}. The strong decay widths of
these states are discussed in Refs. \cite{wei2,zb}.

The possible quantum numbers of $D_{sJ}(2860)$ include
$0^{+}(2^{3}P_{0})$, $1^{-} (1^{3}D_{1})$, $1^- (2 ^3S_1)$, $2^+
(2 ^3P_2)$, $2^+ (1 ^3F_2)$ and $3^{-}(1^{3}D_{3})$. The $2 ^3P_2$
$c\bar s$ state is expected to lie around $(2.95\sim 3.0)$ GeV
while the mass of the $1 ^3F_2$ state will be much higher than
2.86 GeV.

After comparing the theoretical decay widths from the $^{3}P_{0}$
model and decay patterns with the available experimental data, the
authors of Ref. \cite{zb} suggest: (1) $D_{sJ}(2715)$ is probably
the $1^{-}(1^{3}D_{1})$ $c\bar{s}$ state although the
$1^{-}(2^{3}S_{1})$ assignment is not completely excluded; (2)
$D_{sJ}(2860)$ seems unlikely to be the $1^{-}(2^{3}S_{1})$ and
$1^{-}(1^{3}D_{1})$ candidate; (3) $D_{sJ}(2860)$ as either a
$0^{+}(2^{3}P_{0})$ or $3^{-}(1^{3}D_{3})$ $c\bar{s}$ state is
consistent with the experimental data; (4) experimental search of
$D_{sJ}(2860)$ in the channels $D_s\eta$, $DK^{*}$, $D^{*}K$ and
$D_{s}^{*}\eta$ will be crucial to distinguish the above two
possibilities.

%%%%%%%%%%%%%%%%%%%%%%%%%%%%%%%%%%%%%%%%%%%
\subsection{$D_{sj}(2632)$}

\begin{figure}[th]
\scalebox{0.4}{\includegraphics{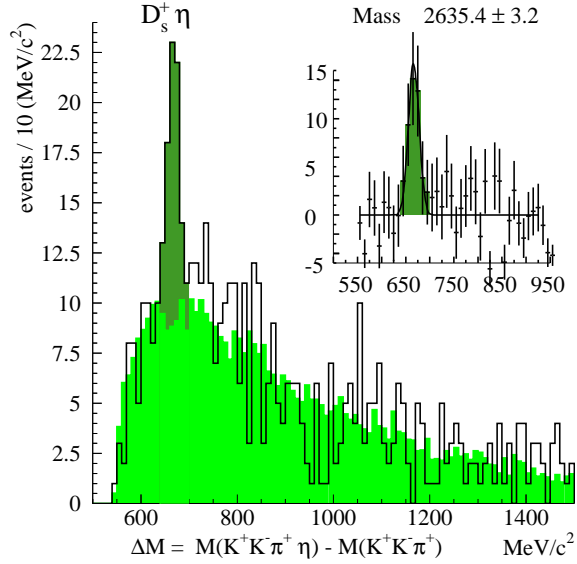}}
%\centerline{\psfig{file=fig17.eps,width=7cm}} \vspace*{8pt}
\caption{$D_{sj}(2632)$ in the $D_s\eta$ spectrum from Ref.
\cite{selex}.\label{figure17}}
\end{figure}

SELEX Collaboration observed an exotic charm-strange meson
$D_{sJ}(2632)$ (Fig. \ref{figure17}).  Its decay width is very
narrow, $\Gamma < 17$ MeV at $90\%$ C.L.  The decay channels are
$D_s \eta$ and $D^0K^+$ with the unusual relative branching ratio
\cite{selex}:
\begin{equation} {\Gamma(D^0K^+) \over
\Gamma(D_s\eta)}=0.16\pm 0.06\; .
\end{equation}
This state lies 274 MeV above $D^0K^+$ threshold and 116 MeV above
$D_s \eta$ threshold. One would naively expect its strong decay
width to be around $(100\sim 200)$ MeV. If SU(3) flavor symmetry
is roughly good, one would expect the relative branching ratio
\cite{lyr1}:
\begin{equation} { \Gamma(D^0K^+)\over
\Gamma(D_s\eta)} \sim 2.3 \; .
\end{equation}

One intriguing possibility is that $D_s(2632)$ is a tetraquark
\cite{maiani4,chen4,nicolescu,chao4,lyr1,gupta}. Then, there
should be other tetraquark partners of $D_{sJ}(2632)$. Moreover,
its narrow width may require that quarks inside $D_{sJ}(2632)$
form tightly bound clusters like diquarks \cite{chao4}.
Tetraquarks with quark content $\bar c\bar q qq$ form four
multiplets: two triplets, one anti-sextet and one 15-plet
\begin{equation}
3\otimes3\otimes\bar{3}\otimes1=3_1\oplus3_2\oplus\bar{6}\oplus15
\; .
\end{equation}
The wave functions of these states can be found in Ref.
\cite{lyr1}. The identification of $D_{sJ}(2632)$ as the $J^P=0^+$
isoscalar member of the ${\bf 15}$ tetraquarks with the quark
content ${1\over 2\sqrt{2}}
(ds\bar{d}+sd\bar{d}+su\bar{u}+us\bar{u}-2ss\bar{s})\bar{c}$ leads
to the relative branching ratio \cite{lyr1}
\begin{equation}
{\Gamma(D_{sJ}^+(2632)\rightarrow {D}^0 K^+)\over
{\Gamma(D_{sJ}^+(2632)\rightarrow D_s^+\eta)}} =0.25\; .
\end{equation}
This decay pattern arises from the SU(3) Clebsch-Gordan
coefficients very naturally.

Another possibility is that $D^+_{sJ}(2632)$ is dominated by
$c\bar ss\bar s$ with $J^P=0^+$ \cite{chen4,gupta}. The $s\bar s$
fluctuates into ${1\over \sqrt{3} \eta_1} -{2\over
\sqrt{6}\eta_8}$. Hence, its decay mode is mainly $D^+_s \eta$.
The final states $D^0K^+, D^+K^0$ are produced through the
annihilation of $s\bar s$ into $u\bar u+d\bar d$ which requires
$\eta_1$ component. Thus this process is OZI suppressed. In this
way the anomalous decay pattern is achieved.

In the diquark correlation configuration \cite{chao4},
$D^+_{sJ}(2632)$ is suggested to be a $(cs)_{3^*}-(\bar s\bar
s)_3$ state where the subscript numbers are color representations.
With the assumption that $(cs)_{3^*}-(\bar s\bar s)_3$ has small
mixing with $(c\bar s)_1-(s\bar s)_1$, one can give a nice
interpretation for the narrow width. The mixing between $s\bar s$
and $u\bar u+d\bar d$ can lead to the unusual branching ratio.

All the above three tetraquark interpretations predict the same
production rates for $D^+K^0, D^0K^+$ final states. A serious
challenge is that SELEX Collaboration didn't find any signal in
the $D^+K^0$ channel \cite{selex}.

A very different tetraquark version $cd\bar d\bar s$ is proposed
in Refs. \cite{maiani4,nicolescu}. Naively, one would expect
$D^0K^+$ decay channel is suppressed while $D^+ K^0$ and $D_s
\eta$ modes are both important. Interestingly, Ref. \cite{maiani4}
invoked the isospin symmetry breaking to explain the relative
branching ratio. It was proposed that the mass eigenstate
$D^+_{sJ}(2632)$ is the mixture between the two flavor eigenstates
$a^+_{c\bar s}$ and $f^+_{c\bar s}$, where $a^+_{c\bar s}$ is the
$I=1$, $I_3=0$ state in $SU(3)_F$ 6 representation and $f^+_{c\bar
s}$ is the $I=0$ state in $\bar{3}_1$. With some special mixing
scheme, the relative ratio is found to be
\begin{equation}
{\Gamma(D^0K^+)\over \Gamma(D^+_s\eta)}=0.16 \;.
\end{equation}
At the same time, the authors of Ref. \cite{maiani4} predicted
\begin{equation}
4< {\Gamma(D^+K^0)\over \Gamma(D^+_s\eta)}<7.6\;,
\end{equation}
\begin{equation}
1.7< {\Gamma(D_s \pi^0)\over \Gamma(D^+_s\eta)}<6.5\;.
\end{equation}

The possibility of $D_{sJ}(2632)$ being a heavy hybrid meson was
discussed in Ref. \cite{lyr2}. More traditional schemes also
exist. With a many-coupled-channel model for non-exotic
meson-meson scattering, Beveren and Rupp noticed that there exists
a resonance at 2.61 GeV with a width of about 8 MeV and nearly the
same decay pattern as required by SELEX experiment \cite{rupp4}.
Within their scheme, several additional states were generated
together with $D_{sJ}(2632)$.

$D_{sJ}(2632)$ was also proposed as the first radial excitation of
$D_s(2112)$ with $J^P=1^-$ \cite{chao4,barnes4}. The nodal
structure of the radial wave function of $D_s(2632)$ may lead to
the narrow width while different decay momentum in two channels
lead to anomalous decay pattern. In Ref. \cite{zzx}, a
quantitative analysis of $D^{+}_{sJ}(2632)$ as the first radial
excitation of $D^{*}_{s}(2112)$ was performed using the
instantaneous Bethe-Salpeter equation. The mass is $2658\pm 15$
MeV. Its width is around 21.7 MeV from PCAC and low energy
theorem, slightly above SELEX's upper bound. The $S-D$ wave mixing
mechanism is explored in order to explain the special decay
pattern.

Unfortunately BABAR, CLEO and FOCUS reported negative results in
their search of $D_{sJ}(2632)$ \cite{nega}, although the
production ratios differ between $e^+e^-$ annihilation and
hadro-production experiments. Very probably $D_{sJ}(2632)$ was an
experimental artifact.

%%%%%%%%%%%%%%%%%%%%%%%%%%%%%%%%%%%%%%%%%%%%%%%%%%%%%%%%
\section{Charmonium or charmonium-like states }\label{sec4}
%%%%%%%%%%%%%%%%%%%%%%%%%%%%%%%%%%%%%%%%%%%%%%%%%%%%%%%%

There has been important progress in the charmonium spectroscopy
in the past few years. Several previously "missing" states were
observed, which are expected in the quark model. Quite a few
unexpected states are discovered experimentally, seriously
challenging the quark model. These new states were named
alphabetically as XYZ etc. Aspects of these XYZ states have been
reviewed in literature, for example in Refs.
\cite{review8a,review8b,review8c,review8d,review8e,review8f}.
Interested readers may also consult these papers.

%%%%%%%%%%%%%%%%%%%%
\subsection{X(3872)}

\subsubsection{The discovery of X(3872)}

\begin{figure}[th]
\scalebox{0.5}{\includegraphics[angle=270]{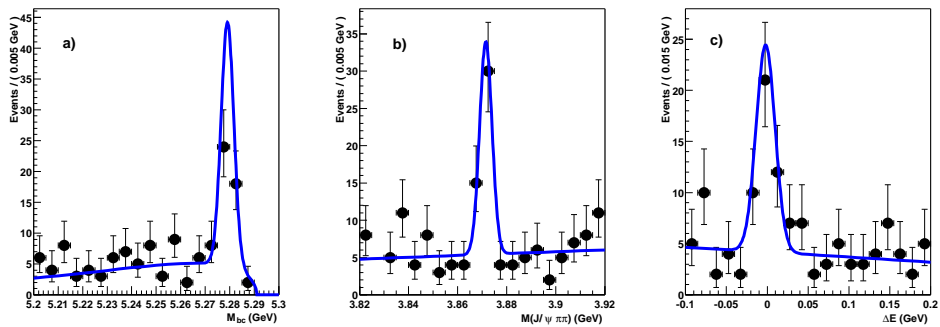}}
%\centerline{\psfig{file=fig18.eps,width=3cm}} \vspace*{8pt}
\caption{X(3872) in the $J/\psi \pi^+\pi^-$ channel from Ref.
\cite{belle8a}.\label{figure18}}
\end{figure}

The charmonium-like state X(3872) was discovered by Belle
collaboration in the $J/\psi \pi^+\pi^-$ channel (Fig.
\ref{figure18}) in the B meson decays \cite{belle8a}. Its
existence was confirmed in the proton anti-proton collisions
shortly after after its discovery and its production property was
found to be similar to that of $\psi^\prime$ by CDF and D0
collaborations \cite{cdf8a,d08}. Later Babar collaboration also
observed X(3872) from B decays \cite{babar8a,babar8b}. From these
measurements its averaged mass is $3871.2\pm 0.5$ MeV \cite{pdg}.
This state is extremely narrow. Its width is less than 2.3 MeV,
consistent with the detector resolution. X(3872) sits almost
exactly on the $D^0{\bar D}^{0\ast}$ threshold and lies very close
to $\rho J/\psi$, $\omega J/\psi$ and $D^+  D^{-\ast}$ threshold.

Both Belle and Babar collaborations reported the radiative decay
mode $X(3872)\to \gamma J/\psi$ \cite{belle8b,babar8c}, which
establishes its charge conjugation parity is even. CDF
collaboration found that the dipion spectrum peaked around the rho
mass \cite{cdf8b}. In other words, the discovery decay mode
violates isospin and G-parity through the decay channel:
$X(3872)\to \rho J/\psi \to J/\psi \pi^+ \pi^-$. Belle
collaboration reported the triple pion decay mode with the triple
pion spectrum peaking around a virtual omega meson \cite{belle8b}.
Such a decay conserves isospin and G-Parity and seems to occur
through the channel: $X(3872)\to \omega J/\psi \to J/\psi \pi^+
\pi^-\pi^0$.

Belle collaboration studied possible quantum number of X(3872)
\cite{belle8c}. They ruled out the $0^{++}$ and $0^{-+}$
possibilities from the angular correlations between final state
particles in $X(3872)\to \pi^+\pi^- J/\psi$ decays. For the
angular momentum between the dipion and $J/\psi$, S-wave is
favored over P-wave from their analysis of the shape of the
$\pi^+\pi^-$ mass distribution near its upper kinematic limit.
Hence both $1^{-+}$ and $2^{-+}$ assignments are strongly
disfavored for X(3872). Belle's analysis strongly favors $1^{++}$
for X(3872), although the $2^{++}$ possibility is not ruled out.
Recently CDF collaboration found that only $1^{++}$ and $2^{-+}$
assignments are consistent with their data from the analysis of
angular distributions and correlations of X(3872) \cite{cdf8c}.
Combining the above experiments one concludes the $J^{PC}$ of
X(3872) is $1^{++}$.

\subsubsection{Theoretical interpretations of X(3872)}

The $2 ^3P_1$ $c\bar c$ state ($\chi_{c1}^\prime$) lies $50\sim
200$ MeV above X(3872) from various quark model calculation. A
charmonium state does not carry isospin, hence does not decay into
$\rho J/\psi$ final states easily. It was quite difficult to find
a suitable position for X(3872) if it was assigned as a $c\bar c$
state \cite{barnes8-qm,eichten8}.

Up to now, theoretical interpretations of X(3872) include a
molecular state \cite{close8,v8,wong8,swanson8,tornqvist8},
$1^{++}$ cusp \cite{bugg8} or S-wave threshold effect
\cite{swave8} due to the $D^0\bar D^{0\ast}$ threshold, hybrid
charmonium \cite{lba8}, diquark anti-diquark bound state
\cite{miani8}, tetraquark state
\cite{richard8,ebert8,vijande8,cui8,narison8,tw8}.

The hybrid charmonium is not expected to lie so low as around 3872
MeV in the flux tube model or from the lattice simulation
\cite{hy8}. Babar collaboration found no charged partners of
X(3872), which were predicted in the diquark anti-diquark bound
state model \cite{miani8}. Its extremely narrow width and lack of
member states in the same multiplet cast doubt on the various
versions of tetraquark hypothesis
\cite{richard8,ebert8,vijande8,cui8,narison8}.

Among the above theoretical schemes, the molecular picture has
been the most popular one. In Ref. \cite{swanson8} Swanson
proposed that (1) X(3872) was mainly a $D^0\bar D^{0\ast}$
molecule bound by both the quark exchange and pion exchange;
(2)its wave function also contains some small but important
admixture of $\rho J/\psi$ and $\omega J/\psi$ components. This
molecular picture explains the proximity of X(3872) to the
$D^0\bar D^{0\ast}$ threshold and the isospin violating $\rho
J/\psi$ decay mode in a very natural way. Moreover this picture
predicted the decay width of the $\pi^+\pi^-\pi^0 J/\psi$ mode is
the same as that of $\rho J/\psi$. This was a remarkable
prediction, confirmed by Belle collaboration later \cite{belle8b}.
So far, the molecular picture was very successful.

There was another nice feature of this model. If X(3872) is a
loosely bound S-wave molecular state, the branching ratio for
$B^0\to X(3872) K^0$ was suppressed by more than one order of
magnitude compared to that for $B^+\to X(3872)K^+$ as shown quite
rigorously by Braaten and Kusunoki in Ref. \cite{bra8}.

\subsubsection{Experimental evidence against the molecular
assignment}

But as more experimental data was released, evidence against the
molecular picture is gradually accumulating. In order to make
comparison with experiments, let's quote the decay widths of
several typical decay modes from Ref. \cite{swanson8}. These modes
had been measured experimentally. With a typical binding energy of
1 MeV for X(3872), the theoretical decay widths of $D^0\bar D^0
\pi^0$, $\pi^+\pi^- J/\psi$, $\pi^+\pi^-\pi^0 J/\psi$ and $\gamma
J/\psi$ are 66 keV, 1215 keV, 820 keV and 8 keV respectively.

The radiative decay mode is clean and ideal to test the model. Two
experiments measured this ratio. The value from Belle
collaboration is \cite{belle8b}
\begin{equation}
{B\left(  X(3872)\to \gamma J/\psi \right)\over B \left(
X(3872)\to \pi^+\pi^- J/\psi \right) } = 0.14\pm 0.05
\end{equation}
and that from Babar collaboration is \cite{babar8c}
\begin{equation}
{B\left(  X(3872)\to \gamma J/\psi \right)\over B \left(
X(3872)\to \pi^+\pi^- J/\psi \right) } \approx 0.25
\end{equation}
while the theoretical prediction from the molecular model is
0.007.

\begin{figure}[th]
\scalebox{0.7}{\includegraphics{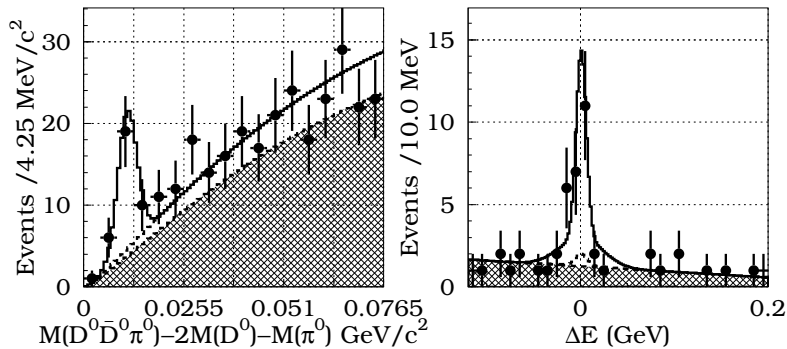}}
%\centerline{\psfig{file=fig19.eps,width=7cm}} \vspace*{8pt}
\caption{The near-threshold enhancement in the $D^0\bar D^0\pi^0$
system from Ref. \cite{belle8d}.\label{figure19}}
\end{figure}

Recently Belle collaboration reported a near-threshold enhancement
(Fig. \ref{figure19}) in the $D^0\bar D^0\pi^0$ system with a mass
$3875.4\pm 0.7 ^{+1.2}_{-2.0}$ MeV \cite{belle8d}. This state can
probably be identified as X(3872). Since it lies 3 MeV above the
$D^0\bar D^{0\ast}$ threshold, it is very awkward to treat it as a
$D^0\bar D^{0\ast}$ molecule.

From the same experiment, we have
\begin{equation}
{B\left( X(3872)\to D^0\bar D^0\pi^0 \right)\over B\left(
X(3872)\to \pi^+\pi^- J/\psi \right) } = 9.4^{+3.6}_{-4.3}
\end{equation}
while the theoretical prediction is 0.054. From Table I in Ref.
\cite{belle8d}, we have
\begin{equation}
{B\left( B^0\to X(3872)K^0 \right)\over B \left( B^+\to X(3872)K^+
\right) } \approx 1.62
\end{equation}
while the theoretical prediction is less than 0.1.

\subsubsection{Is X(3872) still a $1^{++}$ charmonium?}

The similar production properties of X(3872) as $\psi^\prime$ in
the proton anti-proton annihilation and all the above
discrepancies unveil that a very large component of X(3872) or its
short-distance core is still $c\bar c$ or $\chi^\prime_{c1}$, as
also noted in Refs. \cite{chao8,suzuki8}. As $\chi^\prime_{c1}$,
the dominant decay mode is certainly $D^0\bar D^{\ast 0}$ or
$D^0\bar D^0 \pi^0$ at the present mass. Then comes the
hidden-charm decay modes. Its radiative decay $X(3872)\to \gamma
J/\psi$ is also important. As a $1^{++}$ charmonium, its
production properties behave like $\psi^\prime$ naturally.

Previously there were three obstacles with the conventional
charmonium assignment, which can be overcome. The first one is the
isospin violating decay $\rho J/\psi$. The rho-like behavior of
the dipion mass spectrum may result from a dominant decay to the
isospin conserving $\omega J/\psi$ slightly off omega mass shell
and the small isospin breaking rho-omega mixing \cite{suzuki8}.
The enhancement due to the three-body phase space in the
$\pi^+\pi^-J/\psi$ mode compared to the four-body phase space in
the $\pi^+\pi^- \pi^0J/\psi$ mode may compensate the suppression
from the isospin violation to some extent, thus leading to
approximately the same widths. In this case the X(3872) could be
an isoscalar state as expected in a charmonium interpretation.

In Ref. \cite{fsi8} whether the final state interaction effect
plays a significant role in the large hidden charm decay width of
X(3872) is investigated under the assumption that X(3872) is a
candidate of $\chi_{c1}^\prime$. X(3872) decays into $J/\psi \rho$
through exchanging a charmed meson between the $D\bar D^\ast$
intermediate states. In Ref. \cite{fsi8} the authors kept only the
neutral $D\bar D^\ast$ pair and found the absorptive contribution
to $X(3872)\to J/\psi \rho$ is small. Recently the re-scattering
mechanism was reexamined in Ref. \cite{meng8}. Both the charged
and neutral $D\bar D^\ast$ intermediate states were included. The
dispersive contribution was calculated through the dispersion
relation. With the isospin violation in the kinematical factors,
the different Breit-Wigner distribution of the $\rho$ and $\omega$
mesons, and suitable parameters, the authors reproduced the
experimental relative branching ratio between the $J/\psi \rho$
and $J/\psi \omega$ decay modes \cite{meng8}.

The second obstacle is the low mass of X(3872), which may be
ascribed to the large S-wave coupled channel effects, exactly the
same as in the case of $D_{sj}(2317)$. Recall that the quark model
also overestimates the mass of $D_{sj}(2317)$ by 160 MeV. It's
quite plausible that the accuracy of the quark model description
of the hadron spectrum above strong decay threshold is around 100
MeV, especially in presence of the nearby S-wave open channels. In
fact, a recent lattice simulation with improved gauge and Wilson
fermion actions by China Lattice collaboration suggests the mass
of the first excited state of $1^{++}$ charmonium is 3.853(57) GeV
\cite{liuchuan}, consistent with X(3872). The authors also
observed a node structure in the Bethe-Salpeter wave-function of
the $1^{++}$ state, which is characteristic of the radial
excitation.

The third obstacle is the extremely narrow width of X(3872). With
the common $^3P_0$ strong decay model, the total width of X(3872)
is several tens MeV even if it lies only several MeV above the
$D^0\bar D^{0\ast}$ threshold, which is far greater than the
present upper bound. One possible solution is that the strong
decay parameter $\gamma$ in the  $^3P_0$ model decreases near
threshold \cite{chao8a}. The recent analysis in Ref. \cite{meng8}
indicates the total width of X(3872) is about $1\sim 2$ MeV,
consistent with the present upper bound.

%%%%%%%%%%%%%%%%%%%%%%%%%%%%%%%%%
\subsection{Y(4260), Y(4385)}

\begin{figure}[th]
\scalebox{0.5}{\includegraphics{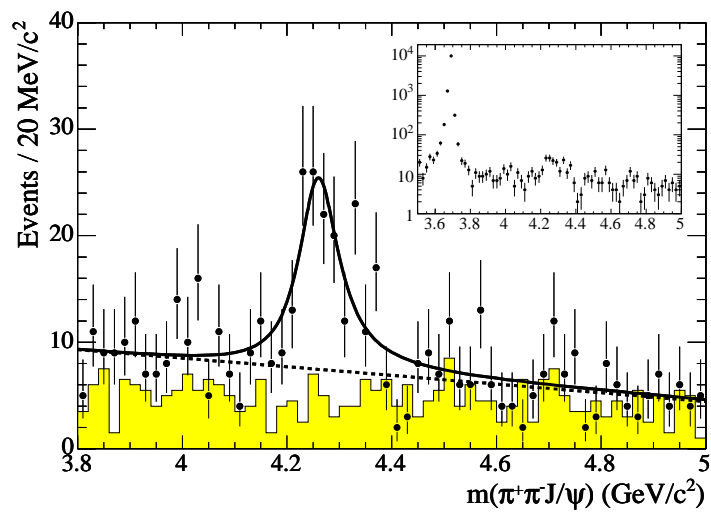}}
%\centerline{\psfig{file=fig20.eps,width=7cm}} \vspace*{8pt}
\caption{Y(4260) in the $\pi^+\pi^- J/\psi$ channel from Ref.
\cite{babar7}.\label{figure20}}
\end{figure}

BABAR collaboration observed a charmonium state around 4.26 GeV in
the $\pi^+\pi^- J/\psi$ channel (Fig. \ref{figure20})
\cite{babar7}. Its decay width is around $(50\sim 90)$ MeV. This
state lies far above threshold but still has a not-so-broad width.
Since this resonance is found in the $e^+e^-$ annihilation through
initial state radiation (ISR), its spin-parity is known
$J^{PC}=1^{--}$. For comparison, R distribution dips around 4.26
GeV \cite{pdg}.

CLEO Collaboration confirmed this state in the $\pi^+\pi^- J/\psi$
channel at 11$\sigma$ significance \cite{cleo7}. Two additional
decay channels $\pi^0\pi^0 J/\psi$ and $K^+ K^- J/\psi$ were
reported at $5.1 \sigma$ and $3.7 \sigma$ significance
respectively. The $e^+e^-$ cross-sections at $s = 4.26$ GeV was
measured to be $\sigma(\pi^+ \pi^- J/\psi) = 58^{+12}_{-10}\pm 4$
pb, $\sigma(\pi^0 \pi^0 J/\psi) = 23^{+12}_{-8} \pm 1$ pb, and
$\sigma(K^+ K^- J/\psi) = 9^{+9}_{-5}\pm 1$ pb. Later, several
other experiments confirmed this state \cite{cleo7a,belle7}. It is
interesting to note that there is no indication for Y(4260) in
other ISR-produced final states. There was also some evidence of
Y(4260) from B decays \cite{babar7c}. For comparison, the central
values of the extracted mass and width of Y(4260) were collected
in Table \ref{tab-y4260}.

\begin{table}[h]
\label{tab-y4260}
\begin{center}
\begin{tabular}{|c|c|c|c|c|}\hline
&Babar&CLEO-c&CLEO III&Belle\\
Events&125&50&14&165\\
Mass&4259&4260&4283&4295\\
Width&88&&70&133\\ \hline
\end{tabular}
\end{center}
\caption{\baselineskip 15pt The central values of the extracted
mass and width of Y(4260) from various experimental measurements.}
\end{table}

According to the PDG assignment of the $1^{--}$ charmonium, there
are four S-wave states $J/\psi$, $\psi(3686)$, $\psi(4040)$,
$\psi(4415)$ and two D-wave states $\psi(3770)$, $\psi(4160)$.
Naively one would expect the $3^3D_1$ state above 4.4 GeV. In
other words, there is no suitable place for Y(4260) in the quark
model spectrum. The discovery of Y(4260) indicates the
overpopulation of the spectrum if PDG classification of the
observed $1^{--}$ charmonium is correct.

A second remarkable feature of Y(4260) is its large hidden-charm
decay width. From BES's measurement of R distribution, the
leptonic width of Y(4260) was extracted \cite{mo}
\begin{equation}
\Gamma[Y(4260)\to e^+ e^-] < 240  \mbox{eV}\; .
\end{equation}
Together with its total width and the measured product
\begin{equation}
\Gamma[Y(4260)\to e^+ e^-] B [Y(4260)\to J\psi \pi^+\pi^-]=5
\mbox{eV}\; ,
\end{equation}
its hidden-charm decay width is found to be:
\begin{equation}
\Gamma[Y(4260)\to J\psi \pi^+\pi^-]>1.8 \mbox{MeV}\; .
\end{equation}
In contrast, the pionic decay width of $\psi^{\prime\prime}$ is
only 50 keV \cite{pdg}. If Y(4260) is a conventional charmonium
state, one might expect comparable hidden-charm width around 200
keV after taking into account the phase space effect. Moreover,
similar dipion transitions from the nearby $\psi(4040)$ or
$\psi(4160)$ states were not observed after careful experimental
scan.

This intriguing state has triggered heated speculations of its
underlying structure. Various models were proposed including a
meson-meson molecular state bound by meson exchange
\cite{liu7,yuan7}, baryonium state composed of two colored
$\Lambda_c$-$\bar \Lambda_c$ clusters \cite{qiao7}, S-wave
threshold effect \cite{jlr7}, a coupled-channel signal \cite{be7},
the first orbital excitation of a diquark-antidiquark state with
the quark content $cs\bar c\bar s$ which decays dominantly into
$D_s\bar D_s$ \cite{miani7}, or a $4S$ charmonium with S-D
interference to explain the dip in the R distribution \cite{4s}.

A recent calculation of the charmonium spectrum above charm
threshold strongly disfavors the charmonium assignment of Y(4260)
in the framework of the Cornell coupled-channel model for the
coupling of $c\bar c$ levels to two-meson states \cite{eichten7}.
It's very interesting to recall that Quigg and Rosner predicted
one $1^{--}$ charmonium state at 4233 MeV using the logarithmic
potential thirty years ago, which was identified as the 4S state
\cite{rosner7}. In order to study possible effects of color
screening and large string tension in heavy quarkonium spectra,
Ding, Chao, and Qin also predicted their 4S charmonium state
exactly at 4262 MeV twelve years ago \cite{chao7}! Their potential
is quite simple:
\begin{equation}
V(r)=-{4\over 3}{\alpha_s\over r} + {T\over \mu}(1-e^{-\mu r})
\end{equation}
where $T$ is the string tension and $\mu$ is the screening
parameter. With such a perfect agreement, one may wonder whether
PDG assignment misses one $1^{--}$ charmonium state in the quark
model. Or does the same traditional quark potential hold for
higher states far above strong decay threshold? However, one
serious challenge remains for the conventional quark model
interpretation: how to generate the huge $J/\psi \pi\pi$ decay
width?

The solution seems to lie beyond the conventional quark model. The
mass of Y(4260) does not contradict the scalar tetraquark
hypothesis. If so, it must be produced by the $I=0$ component of
the virtual photon from the ISR experiments. Then the $I=1, I_z=0$
component of the virtual photon should have produced its isovector
partner $Y^\prime (4260)$, which may be searched for in the decay
channel $\pi^+\pi^-\pi^0 J/\psi$ using exactly the same database
from the initial state radiation process. In fact, the
non-observation of $Y^\prime (4260)$ already rejected the
tetraquark hypothesis. Moreover, a tetraquark far above threshold
can fall apart into $D\bar D, D^\ast\bar D$ very easily. Its
not-so-large width also disfavors the tetraquark hypothesis
\cite{zhu7}.

In fact, the promising and feasible interpretation is the hybrid
charmonium interpretation \cite{zhu7,kou7,close7}. At present,
none of the available experimental information is in conflict with
the hybrid charmonium picture. As a $c\bar c G$ state, it is
expected to lie around this mass range. It couples to the virtual
photon weakly and has a small leptonic width, thus explaining the
dip at 4.26 GeV in the R distribution. The gluon field inside the
hybrid splits into light quark pairs quite easily, leading to the
huge hidden-charm decay width. Except for the phase space, such a
process occurs roughly in a flavor-symmetric way.

There were two lattice simulations of Y(4260). With a molecular
operator $\{(\bar q\gamma_5\gamma_i c)(\bar c \gamma_5 q) -(\bar
c\gamma_5\gamma_i q)(\bar q \gamma_5 c) \}$, a resonance signal
was observed with a mass around $4238\pm 31$ MeV, which was
identified as Y(4260) \cite{tw}. In Ref. \cite{luo} the
possibility of Y(4260) as a hybrid charmonium was investigated on
the lattice.

The width of all hidden-charm decay modes of Y(4260) is around
several MeV while its total width is about 100 MeV. In other
words, its main decay channels have not been observed.
Experimentally it will be helpful to search for the $\bar D
D^\ast_1$, $\bar D^\ast D_0^\ast$, $\bar D D_1$, $\bar D_s
D_{sj}(2317)$ modes etc, where $D_0^\ast$ and $D^\ast_1$ are broad
P-wave scalar mesons in the $(0^+, 1^+)$ doublet while $D_1$ is
the narrow $1^+$ state. Some decay channels may occur through the
upper tail of the mass distribution of Y(4260) since it's broad.

\begin{figure}[th]
\scalebox{0.4}{\includegraphics{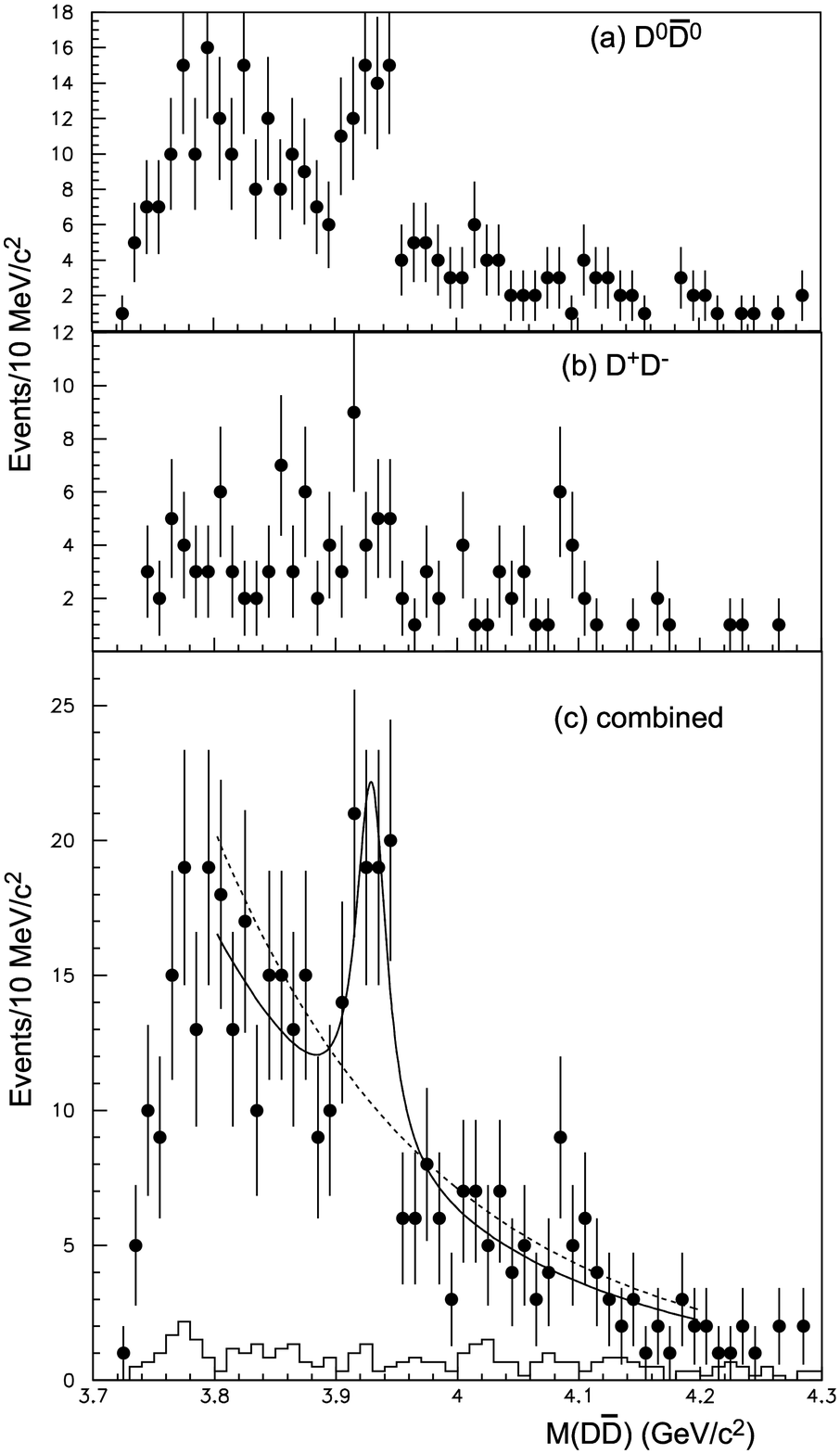}}
%\centerline{\psfig{file=fig21.eps,width=7cm}} \vspace*{8pt}
\caption{$\chi^\prime_{c2}$ in the process $\gamma\gamma \to D\bar
D$ from Ref. \cite{belle9z}.\label{figure21}}
\end{figure}

Babar's recent observation of a new broad signal at 4320 MeV in
the $\psi(2S)\pi^+\pi^-$ channel via the same ISR technique
further complicates the situation \cite{babar7b}. The central
values of the extracted mass from Belle and CLEOIII deviated from
those from Babar and CLEOc, although all these structures are
quite broad. It will be extremely helpful to establish
experimentally whether all these structures including this Y(4320)
arose from one single broad resonance.

%%%%%%%%%%%%%%%%%%%%%%%%%%%%%%%%%%%%%%%%%
\subsection{X(3940), Y(3940), Z(3930)}

\begin{figure}[th]
\scalebox{0.6}{\includegraphics{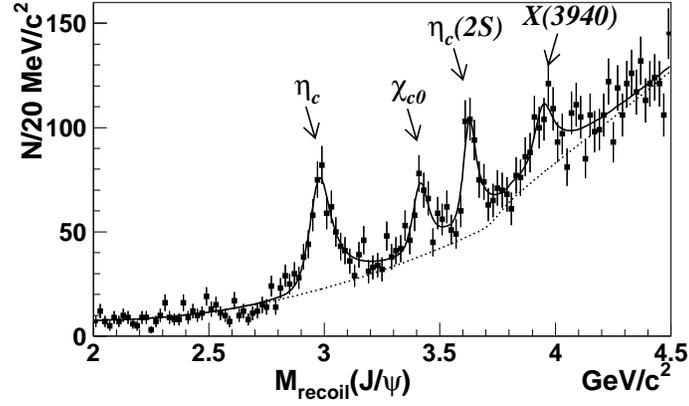}}
%\centerline{\psfig{file=fig22.eps,width=7cm}} \vspace*{8pt}
\caption{X(3940)in the spectrum of masses recoiling from the
$J/\psi$ in the inclusive process $e^+e^- \to J/\psi
+\mbox{anything}$ from Ref. \cite{belle9x}.\label{figure22}}
\end{figure}

Belle collaboration observed a new charmonium state (Fig.
\ref{figure21}) in the process $\gamma\gamma \to D\bar D$
\cite{belle9z}. The production rate and the angular distribution
in the $\gamma\gamma$ center-of-mass frame suggest that this state
is the $2^{++}$ charmonium state $\chi^\prime_{c2}$. Its mass and
width are $M=3931\pm 4\pm 2$ MeV and $\Gamma=20\pm 8 \pm 3$ MeV
respectively. This state matches the theoretical expectation of
the $2^3P_2$ charmonium in the quark model very well
\cite{barnes8-qm}, although its mass is 40-100 MeV below quark
model predictions. In fact, such a discrepancy can be viewed the
typical accuracy of the quark model predictions of the higher
charmonium spectrum above threshold, if one treats
$\chi^\prime_{c2}$ as a benchmark in the comparison between theory
and experiments. As $\chi^\prime_{c2}$, the other main decay mode
$D\bar D^\ast$ awaits to be discovered.

\begin{figure}[th]
\scalebox{0.6}{\includegraphics{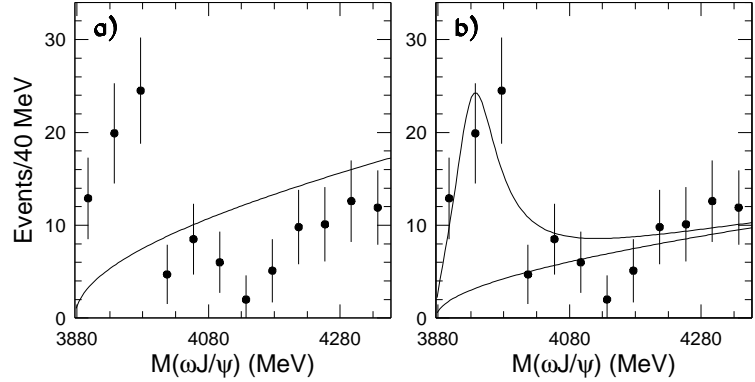}}
%\centerline{\psfig{file=fig23.eps,width=7cm}} \vspace*{8pt}
\caption{Y(3940) in the $\omega J/\psi$ invariant mass
distribution from Ref. \cite{belle9y}.\label{figure23}}
\end{figure}

Around the same mass, Belle collaboration observed another
charmonium-like state X(3940) (Fig. \ref{figure22}) in the
spectrum of masses recoiling from the $J/\psi$ in the inclusive
process $e^+e^- \to J/\psi +\mbox{anything}$ \cite{belle9x}.
X(3940) decays into $\bar D D^\ast$ with a width less than 52 MeV.
$X(3940) \to D\bar D, \omega J/\psi$ modes were not observed. Such
a decay pattern is typical of $\chi^\prime_{c1}$
\cite{barnes8-qm}. If X(3940) is $\chi^\prime_{c1}$, one may
expect a stronger signal of its ground state $\chi_{c1}$. But in
the same reaction, Belle collaboration didn't observe $\chi_{c1}$.
A possible assignment of X(3940) is $\eta_c^{\prime\prime}$,
although the $\eta_c^{\prime\prime}$ mass is predicted to be
around 100 MeV higher in the quark model. Such is discrepancy
should be tolerable if one keeps the $\chi^\prime_{c2}$ case in
mind.

Belle collaboration reported a near-threshold enhancement Y(3940)
in the $\omega J/\psi$ invariant mass distribution (Fig.
\ref{figure23}) for exclusive $B\to K\omega J/\psi$ decays
\cite{belle9y}. Its width is around 92 MeV. The decay width of
$Y(3940)\to \omega J/\psi$ is greater than 7 MeV, which is really
huge since this is a hidden-charm decay and violates isospin!
Neither $D\bar D$ nor $D\bar D^\ast$ decay mode was observed. Such
a decay pattern motivated Belle collaboration to speculate Y(3940)
as a hybrid charmonium. As said before, hybrid charmonium states
are expected to lie higher \cite{lattice}. If it were a hybrid,
the potentially dominant decay modes in the form of $L=0$ and
$L=1$ combination are kinematically forbidden. Then one would
expect $D\bar D$ and $D\bar D^\ast$ to be the main modes. The
priority is to confirm this threshold enhancement in other decay
channels since its main decay modes have not been observed yet.
Right now the possibility of this threshold enhancement being an
artifact is not excluded.

%%%%%%%%%%%%%%%%%%%%%%%%%%%%%%%%%%%%
\section{Conclusion}\label{sec5}
%%%%%%%%%%%%%%%%%%%%%%%%%%%%%%%%%%%%

In this paper new hadron states observed in the past four years
are reviewed. Some of them lie very close to their strong decay
thresholds. Among them, $D_{sj}(2317)$, $D_{sj}(2460)$, X(3872),
Z(3930) and Y(4260) are well established experimentally while som
other states especially those threshold enhancements await
confirmation from other collaborations. To summarize,
\begin{itemize}

\item Z(3930) is $\chi_{c2}^\prime$.

\item Y(4260) is an excellent hybrid charmonium candidate.

\item X(3872) (or its dominant component) is very probably
$\chi_{c1}^\prime$.

\item $D_{sj}(2317)$ and $D_{sj}(2460)$ are $c\bar s$ states.

\item $D_{sj}(2632)$ is probably an experimental artifact.

\end{itemize}

The quark model spectrum for the $0^+, 1^+$ $c\bar s$ and $1^{++}$
$c\bar c$ systems are distorted by nearby open channels. The
strong interference with the S-wave continuum lowers the original
quark model levels significantly. One example is the level
ordering of X(3872) and Z(3930), which indicates the S-wave
couple-channel effect is important for spectrum above or near
strong decay threshold.

The two particle $DK$ continuum state, a $DK$ resonance and a
bound state behave differently on the lattice. A dynamical
simulation of $DK$ scattering with chiral symmetry on the lattice
will be extremely helpful. One may extract their phase shifts and
energy levels. Through the variation of the energy level with the
lattice length $L$, one may get some crucial hints of their
underlying structure. Future lattice simulation of $DK$ and
$D^0\bar D^{\ast 0}$ scattering is strongly called for to shed
light on $D_{sj}(2317)$ and X(3872).

This year BESIII will start taking data in Beijing. Its luminosity
is expected to be 100 times of that for BESII. There are active
experimental programs on hybrid mesons and glueballs at JLAB.
B-factories are collecting data continuously. In the near future,
J-PARC will have a large hadron physics program in Japan. There
will be great advances in hadron spectroscopy and the low energy
sector of QCD with so many experimental facilities running.

%%%%%%%%%%%%%%%%%%%%%%%%%%%%%%%%%%%
\section*{Acknowledgments}
%%%%%%%%%%%%%%%%%%%%%%%%%%%%%%%%%%%

The author thanks Professor Z.-Y. Zhang for suggesting the topic
and all my collaborators for the enjoyable collaborations. Because
of his personal research interest, the author apologizes to those
whose papers and contributions were not mentioned in this work.
This project was supported by the National Natural Science
Foundation of China under Grants 10421503 and 10625521, and
Ministry of Education of China, FANEDD.


\begin{thebibliography}{0}

\bibitem{nakano}T. Nakano et al., Phys. Rev. Lett. 91, 012002 (2003).

\bibitem{lattice}C. Michael, hep-lat/0302001.

\bibitem{witten}E. Witten, Nucl. Phys. B 169, 57 (1979).

\bibitem{zhu0}Shi-Lin Zhu, Int. J. Mod. Phys. A 19, 3439 (2004);
{\bf ibid} A 20, 1548 (2005).

\bibitem{oka0}M. Oka, Prog. Theor. Phys. 112, 1 (2004).

\bibitem{hicks}K. H. Hicks, Prog. Part. Nucl. Phys. 55, 647 (2005).

\bibitem{bes}J. Z. Bai et al, BES Collaboration, Phys. Rev. Lett.
91, 022001 (2003).

\bibitem{gao}C.-S. Gao, Shi-Lin Zhu, Commun. Theor. Phys. 42, 844
(2004).

\bibitem{bes1}M. Ablikim et al., BES Collaboration,
Phys. Rev. Lett. 95, 262001 (2005).

\bibitem{fsi}A.Sirbirtsev et al., hep-ph/0411386.

\bibitem{pdg}W.-M. Yao et al., J. Phys. G 33, 1 (2006).

\bibitem{gao1}Shi-Lin Zhu, C. S. Gao, Commun. Theor. Phys. 46,
291 (2006).

%\cite{rosner}
\bibitem{rosner}
  J.~L.~Rosner,
  %``Low-mass baryon antibaryon enhancements in B decays,''
  Phys.\ Rev.\ D {\bf 68}, 014004 (2003).
  %%CITATION = HEP-PH 0303079;%%

%\cite{Datta:2003iy}
\bibitem{q1}
  A.~Datta and P.~J.~O'Donnell,
  %``A new state of baryonium,''
  Phys.\ Lett.\ B {\bf 567}, 273 (2003).
  %%CITATION = HEP-PH 0306097;%%

%\cite{Zou:2003zn}
\bibitem{q2}
  B.~S.~Zou and H.~C.~Chiang,
  %``One-pion exchange final-state interaction and the p anti-p near  threshold
  %enhancement in J/psi $\to$ gamma p anti-p decays,''
  Phys.\ Rev.\ D {\bf 69}, 034004 (2004).
  %%CITATION = HEP-PH 0309273;%%

%\cite{Kerbikov:2003qg}
\bibitem{q3}
  B.~Kerbikov, A.~Stavinsky and V.~Fedotov,
  %``Low-mass proton antiproton enhancement: Belle and BES results, premises  of
  %LEAR and expectations from CLAS,''
  arXiv:nucl-th/0310060.
  %%CITATION = NUCL-TH 0310060;%%

%\cite{Kerbikov:2004gs}
\bibitem{q4}
  B.~Kerbikov, A.~Stavinsky and V.~Fedotov,
  %``Model-independent view on the low-mass proton antiproton enhancement,''
  Phys.\ Rev.\ C {\bf 69}, 055205 (2004).
  %%CITATION = HEP-PH 0402054;%%

%\cite{Mishustin:2004xa}
\bibitem{q5}
  I.~N.~Mishustin, L.~M.~Satarov, T.~J.~Burvenich, H.~Stoecker and W.~Greiner,
  %``Antibaryons bound in nuclei,''
  Phys.\ Rev.\ C {\bf 71}, 035201 (2005).
  %%CITATION = NUCL-TH 0404026;%%

%\cite{Yan:2004xs}
\bibitem{q6}
  M.~L.~Yan, S.~Li, B.~Wu and B.~Q.~Ma,
  %``Nucleon antinucleon bound states in Skyrmion-type potential,''
  Phys.\ Rev.\ D {\bf 72}, 034027 (2005).
  %%CITATION = HEP-PH 0405087;%%

%\cite{Liu:2004er}
\bibitem{q7}
  X.~A.~Liu, X.~Q.~Zeng, Y.~B.~Ding, X.~Q.~Li, H.~Shen and P.~N.~Shen,
  %``Can the observed enhancement in the mass spectrum of p anti-p in J/psi
  %$\to$ gamma p anti-p be interpreted by a possible p anti-p bound state,''
hep-ph/0406118.
  %%CITATION = HEP-PH 0406118;%%

%\cite{Bugg:2004rk}
\bibitem{q8}
  D.~V.~Bugg,
  %``Reinterpreting several narrow 'resonances' as threshold cusps,''
  Phys.\ Lett.\ B {\bf 598}, 8 (2004).
  %%CITATION = HEP-PH 0406293;%%

%\cite{He:2004ih}
\bibitem{q9}
  X.~G.~He, X.~Q.~Li and J.~P.~Ma,
  %``Parity, charge conjugation and SU(3) constraints on threshold enhancement
  %in J/psi decays into gamma p anti-p and K p anti-Lambda,''
  Phys.\ Rev.\ D {\bf 71}, 014031 (2005).
  %%CITATION = HEP-PH 0407083;%%

%\cite{Chang:2004us}
\bibitem{q10}
  C.~H.~Chang and H.~R.~Pang,
  %``On possible S-wave bound states for a N anti-N system within a constituent
  %quark model,''
  Commun.\ Theor.\ Phys.\  {\bf 43}, 275 (2005).
  %%CITATION = HEP-PH 0407188;%%

\bibitem{q11}A.Sirbirtsev et al., hep-ph/0411386.

%\cite{Loiseau:2004te}
\bibitem{q12}
  B.~Loiseau and S.~Wycech,
  %``Antiproton proton resonant like channels in J/psi $\to$ gamma p anti-p
  %decays,''
  Int.\ J.\ Mod.\ Phys.\ A {\bf 20}, 1990 (2005).
  %%CITATION = HEP-PH 0411218;%%

%\cite{Loiseau:2005cv}
\bibitem{q13}
  B.~Loiseau and S.~Wycech,
  %``Antiproton-proton channels in J/psi decays,''
hep-ph/0501112.
  %%CITATION = HEP-PH 0501112;%%

%\cite{Ding:2005ew}
\bibitem{q14}
  G.~J.~Ding and M.~L.~Yan,
  %``Proton antiproton annihilation in baryonium,''
hep-ph/0502127.
  %%CITATION = HEP-PH 0502127;%%

%\cite{Ding:2005tr}
\bibitem{q15}
  G.~J.~Ding, J.~l.~Ping and M.~L.~Yan,
  %``Baryon antibaryon enhancements in quark models,''
hep-ph/0510013.
  %%CITATION = HEP-PH 0510013;%%

%\cite{Kochelev:2005vd}
\bibitem{koch}
  N.~Kochelev and D.~P.~Min,
  %``X(1835) as the lowest mass pseudoscalar glueball and proton spin problem,''
hep-ph/0508288, hep-ph/0510016.
  %%CITATION = HEP-PH 0508288;%%

%\cite{hxg}
\bibitem{hxg}
  X.~G.~He, X.~Q.~Li, X.~Liu and J.~P.~Ma,
  %``Some properties of the newly observed X(1835) state at BES,''
hep-ph/0509140.
  %%CITATION = HEP-PH 0509140;%%

%\cite{Li:2005vd}
\bibitem{lba}
  B.~A.~Li,
  %``A possible 0-+ glueball candidate X(1835),''
hep-ph/0510093.
  %%CITATION = HEP-PH 0510093;%%

\bibitem{fermi}E. Fermi and C. N. Yang, Phys. Rev. 76, 1739
(1949).
\bibitem{history}J.-M. Richard, Nucl. Phys. B (Proc. Suppl.) 86,
361 (2000).
\bibitem{pr}E. Klempt, F. Bradamante, A. Martin, and J.-M.
Richard, Phys. Rep. 368, 119 (2002) and references herein.

\bibitem{tao}T. Huang, Shi-Lin Zhu, Phys. Rev. D 73, 014023 (2006).

\bibitem{1576-BES}M. Ablikim et al., BES Collaboration,
Phys. Rev. Lett. 97, 142002 (2006).

\bibitem{lx1}X. Liu, B. Zhang, L.-L. Shen, Shi-Lin Zhu,
hep-ph/0701022.

\bibitem{Guo-1576}F.K. Guo and P.N. Shen, Phys. Rev. D 74, 097503
(2006).

\bibitem{Lipkin-1576}M. Karliner and H.J. Lipkin, hep-ph/0607093.

\bibitem{1576-QSR}Z.G. Wang and S.L. Wang, Chin. Phys. Lett. 23, 3208
(2006).

\bibitem{Ding-1576}G.J. Ding and M.L. Yan, Phys. Lett. B 643, 33 (2006).

\bibitem{zhang}A.L. Zhang, T. Huang and T.G. Steele, hep-ph/0612146.

\bibitem{bes3}M. Ablikim et al., BES Collaboration, Phys. Lett. B 607, 243
(2005).

\bibitem{bes4}M. Ablikim et al., BES Collaboration, Phys. Rev. Lett. 96,
162002 (2006).

\bibitem{tetra1}
  B.~A.~Li,
  %``Production of a Q Q anti-Q anti-Q state in J/psi --> gamma omega Phi,''
  Phys.\ Rev.\  D {\bf 74}, 054017 (2006).

\bibitem{tetra2}
  F.~Buccella, H.~Hogaasen, J.~M.~Richard and P.~Sorba,
  %``Chromomagnetism, flavour symmetry breaking and S-wave tetraquarks,''
  Eur.\ Phys.\ J.\  C {\bf 49}, 743 (2007).

\bibitem{tetra3}
  A.~Zhang, T.~Huang and T.~G.~Steele,
  %``Diquark and light four-quark states,''
hep-ph/0612146.

\bibitem{hybrid}
  K.~T.~Chao,
  %``A short note on q anti-q g hybrid assignment for X(1812) --> omega Phi,''
hep-ph/0602190.

\bibitem{hybrid1}
M.S. Chanowitz and S.R. Sharpe, Phys. Lett. B 132, 413 (1983);
Nucl. Phys. B 222, 211 (1983).

\bibitem{glue1}
  P.~Bicudo, S.~R.~Cotanch, F.~J.~Llanes-Estrada and D.~G.~Robertson,
  %``The BES f_0(1810): A New Glueball Candidate,''
hep-ph/0602172.

\bibitem{glue2}
  D.~V.~Bugg,
  %``A glueball component in f0(1790),''
hep-ph/0603018.

\bibitem{mix1}
  X.~G.~He, X.~Q.~Li, X.~Liu and X.~Q.~Zeng,
  %``X(1812) in quarkonia-glueball-hybrid mixing scheme,''
  Phys.\ Rev.\  D {\bf 73}, 114026 (2006).

\bibitem{threshold}
  J.~L.~Rosner,
  %``Effects of S-wave thresholds,''
  Phys.\ Rev.\  D {\bf 74}, 076006 (2006).

\bibitem{zhao-fsi}
  Q.~Zhao and B.~S.~Zou,
  %``On the near-threshold enhancement in J/psi --> gamma X with X --> omega
  %Phi,''
  Phys.\ Rev.\  D {\bf 74}, 114025 (2006).

\bibitem{cohen}T. D. Cohen, Phys. Lett. B 427, 348 (1998).

\bibitem{flux1}F. E. Close, P.R. Page, Nucl. Phys. B 443, 233 (1995);
F. E. Close, S. Godfrey, Phys. Lett. B 574, 210 (2003).

\bibitem{flux2}P.R. Page, E.S. Swanson and A.P. Szczepaniak. Phys. Rev.
{\bf D59}, 034016 (1999).

\bibitem{zhu-qcd}Shi-Lin Zhu, Phys. Rev. D 60, 014008 (1999).

\bibitem{bali}G. S. Bali, Phys. Rev. D 71, 114513 (2005).

\bibitem{etapi1}S.U. Chung, et al., Phys. Rev. D60, 092001-1 (1999), and
references therein.

\bibitem{etapi2}A. Abele et al., Phys. Lett. B446, 349 (1999).

\bibitem{rhopi}G. S. Adams et al. Phys. Rev. Lett. 81, 5760 (1998); S. U.
Chung et al., Phys. Rev. D65, 072001 (2002).

\bibitem{etappi}E.I. Ivanov, et al., Phys. Rev. Lett. 86, 3977 (2001).

\bibitem{f1pi}J. Kuhn, et al., Phys. Lett. B 595, 109 (2004).

\bibitem{b1pi}M. Lu et al., Phys. Rev. Lett. 94, 032002 (2005).

\bibitem{challenge1}A. P. Szczepaniak, M. Swat, A. R. Dzierba, Phys. Rev.
Lett. 91, 092002 (2003); A.R. Dzierba et al., Phys. Rev. D 67,
094015 (2003).

\bibitem{confirm}G. S. Adams et al., hep-ex/0612062.

\bibitem{challenge2}A. R. Dzierba et al., Phys. Rev. D 73, 072001
(2006).

%%%%%%%%%Y(2175)
\bibitem{babar10}B. Aubert et al., Phys. Rev. D 74, 091103 (2006).

\bibitem{wzg10}Z. G. Wang, hep-ph/0610171.

\bibitem{ding10a}G. J. Ding, M. L. Yan, hep-ph/0611319.

\bibitem{barnes10}T. Barnes, N. Black and P. R. Page,
Phys. Rev. D 68, 054014 (2003).

\bibitem{ding10b}G. J. Ding, M. L. Yan, hep-ph/0701047.

%%%%%%%%%%%%p anti-Lambda
\bibitem{bes11}M. Ablikim et al., Phys. Rev. Lett. 93, 112002 (2004).
\bibitem{belle11}M. Z. Wang et al., Phys. Lett. B 617, 141 (2005).

\bibitem{he11}X. G. He, X. Q. Li, J. P. Ma, Phys. Rev. D 71, 014031 (2005).
\bibitem{yuan11}C. Z. Yuan, X. H. Mo, P. Wang, Phys. Lett. B 626, 95 (2005).
\bibitem{ding11}G. J. Ding, J. L. Ping, M. L. Yan,
Phys. Rev. D 74, 014029 (2006).
\bibitem{wu11}F. Q. Wu, B. S. Zou, Phys. Rev. D 73, 114008 (2006).

%%%%%%%%%%%%
\bibitem{bes12}M. Ablikim et al., Phys. Rev. Lett. 97, 062001 (2006).


%%%%%%%%%%%%%Charmed mesons
\bibitem{HQET}B. Grinstein, Nucl. Phys. B339, 253 (1990); E. Eichten and B.
Hill, Phys. Lett. B 234, 511 (1990); A. F. Falk, H. Georgi, B.
Grinstein, and M. B. Wise, Nucl. Phys. B343, 1 (1990).

\bibitem{focus1}K. Abe et al., Phys. Rev. D 69, 112002 (2004).

\bibitem{belle1}J. M. Link et al., Phys. Lett. B 586, 11 (2004).

\bibitem{pol}V. Dmitrasinovic, Phys. Rev. Lett. 94, 162002 (2005).

\bibitem{babar1} Babar Collaboration, B. Aubert et al., Phys. Rev. Lett. 90, 242001 (2003).
\bibitem{cleo}   CLEO Collaboration, D. Besson et al., Phys. Rev. D 68, 032002 (2003).
\bibitem{belle1b} Belle Collaboration, P. Krokovny et al., Phys. Rev. Lett. 91, 262002 (2003).
\bibitem{belle2} Belle Collaboration, Y.Mikami et al., Phys. Rev. Lett. 92, 012002 (2004).
\bibitem{belle3} Belle Collaboration, A. Drutskoy  et al., Phys. Rev. Lett. 96, 061802 (2005).
\bibitem{belle4} Belle Collaboration, P. Krokovny et al., hep-ex/0310053.
\bibitem{focus}  FOCUS Collabortation, E. W. Vaandering, hep-ex/0406044.
\bibitem{babar2} Babar Collaboration, B. Aubert et al., Phys. Rev. Lett. 93, 181801 (2004).
\bibitem{babar3} Babar Collaboration, B. Aubert et al., Phys. Rev. D. 69, 031101 (2004).
\bibitem{babar4} Babar Collaboration, G. Calderini et al., hep-ex/0405081.
\bibitem{babar5} Babar Collaboration, B. Aubert et al., hep-ex/0408067.

\bibitem{qm}S. Godfrey and R. Kokoski, Phys. Rev.
D 43, 1679 (1991); S. Godfrey and N. Isgur, Phys. Rev.  D 32, 189
(1985); M. Di Pierro and E. Eichten, Phys. Rev. D 64, 114004
(2001).

\bibitem{jin}Y.-B. Dai, C.-S. Huang and H.-Y. Jin, Phys. Lett. B 331, 174 (1994).

\bibitem{bali2}G. S. Bali, Phys. Rev. D 68, 071501 (2003).

\bibitem{dougall} A. Dougall, R.D. Kenway, C.M. Maynard,
and C. McNeile, Phys. Lett. B 569, 41 (2003).

\bibitem{soni}H. W. Lin, S. Ohta, A. Soni, and N. Yamada,
hep-lat/0607035.

\bibitem{barnes}T. Barnes, F. E. Close, H. J. Lipkin, Phys.
Rev. D 68, 054006 (2003).

\bibitem{rupp}
E. van Beveren and G. Rupp, Phys. Rev. Lett. 91, 012003 (2003);
Euro. Phys. J. C 32, 493 (2004).

\bibitem{bardeen}W. A. Bardeen, E. J. Eichten, C. T. Hill,
Phys. Rev. D 68, 054024 (2003).

\bibitem{cheng}H. Y. Cheng, W. S. Hou, Phys. Lett. B 566, 193
(2003).

\bibitem{kt}K. Terasaki, Phys. Rev. D68, 011501 (2003).

\bibitem{vijande}J. Vijande, F. Fernandez, A. Valcarce,
Phys. Rev. D 73, 034002 (2006).

\bibitem{zzy}H. X. Zhang, W. L. Wang, Y. B. Dai, and Z. Y. Zhang,
hep-ph/0607207.

\bibitem{godfrey}S. Godfrey, Phys. Lett. B568, 254 (2003).

\bibitem{col1}P. Colangelo and F. De Fazio, Phys.
Lett. B570, 180 (2003).

\bibitem{dai1}Y. B. Dai, C. S. Huang, C. Liu, Shi-Lin Zhu,
Phys. Rev. D 68, 114011 (2003).

\bibitem{wise}P. L. Cho and M. B. Wise, Phys. Rev. D 49,
6228 (1994).

\bibitem{col2}P. Colangelo, F. De Fazio and A. Ozpineci,
Phys. Rev. D 72, 074004 (2005).

\bibitem{wei1}W. Wei, P.-Z. Huang, and Shi-Lin Zhu, Phys.
Rev. D 73, 034004 (2006).

\bibitem{lu}J. Lu, W.-Z. Deng, X.-L. Chen, and Shi-Lin Zhu,
Phys. Rev. D 73, 054012 (2006).

\bibitem{faya}Fayyazuddin and Riazuddin, Phys. Rev. D 69, 114008 (2004).

\bibitem{ishida}S. Ishida et al., hep-ph/0310061.

\bibitem{colangelo91}P. Colangelo, G. Nardulli,
A. A. Ovchinnikov , N. Paver, Phys. Lett. B 269, 201 (1991).

\bibitem{haya}A. Hayashigaki, K. Terasaki, hep-ph/0411285.

\bibitem{narison}S. Narison, Phys. Lett. B 605, 319 (2005).

\bibitem{colangelo95}P. Colangelo, F. De Fazio, G. Nardulli,
N. Di Bartolomeo, Raoul Gatto, Phys. Rev. D52, 6422 (1995).

\bibitem{zhuzhu}S.-L. Zhu, Y.-B. Dai, Mod. Phys. Lett. A 14, 2367
(1999).

\bibitem{shifman}B. Blok, M. Shifman, N. Uraltsev, Nucl. Phys.
B494,247(1997).

%\bibitem{dai2}Y. B. Dai, Shi-Lin Zhu, Y. B. Zuo, hep-ph/0610327.
%\bibitem{dai3}Y. B. Dai, Shi-Lin Zhu, hep-ph/0611318.

\bibitem{babar-high}B. Aubert et al. (BABAR Collaboration),
Phys. Rev. Lett. 97, 222001 (2006).

\bibitem{belle-high}K. Abe et al. (Belle Collaboration), hep-ex/0608031.

\bibitem{rupp3}E.V. Beveren and G. Rupp, Phys. Rev. Lett. 97, 202001 (2006).

\bibitem{close3}F.E. Close, C.E. Thomas, O. Lakhina and E.S.
Swanson, hep-ph/0608139.

\bibitem{colangelo3}P. Colangelo, F.D. Fazio and S. Nicotri,
Phys. Lett. B 642, 48 (2006).

\bibitem{isgur3}S. Godfrey and N. Isgur, Phys. Rev. D 32, 189 (1985).

\bibitem{nowak}M. A. Nowak, M. Rho, I. Zahed, Acta Phys. Polon. B
35, 2377 (2004).

\bibitem{wei2}W. Wei, X. Liu, Shi-Lin Zhu, Phys. Rev. D 75,
014013 (2007).

\bibitem{zb}B. Zhang, X. Liu, W. Z. Deng, Shi-Lin Zhu, Euro. Phys.
J. C (2007), hep-ph/0609013.

\bibitem{selex}A.V. Evdokimov et al., Phys. Rev. Lett. 93, 242001
(2004).

\bibitem{lyr1}Y. R. Liu, Shi-Lin Zhu, Y. B. dai, C. Liu,
Phys. Rev. D 70, 094009 (2004).

\bibitem{maiani4}L. Maiani et al., Phys. Rev. D 70, 054009 (2004).

\bibitem{chen4}Y.-Q. Chen and X.-Q. Li, Phys. Rev. Lett. 93, 232001 (2004).

\bibitem{nicolescu}B. Nicolescu and J. Melo, hep-ph/0407088.

\bibitem{chao4}K.-T. Chao, Phys. Lett. B 599, 43 (2004).

\bibitem{barnes4}T. Barnes et al., Phys. Lett. B 600, 223 (2004).

\bibitem{gupta}V. Gupta, Int. J. Mod. Phys. A 20, 5891 (2005).

\bibitem{lyr2}Y. B. Dai, C. Liu, Y. R. Liu, Shi-Lin Zhu,
JHEP 0411, 043 (2004).

\bibitem{rupp4}E. van Beveren, G. Rupp, Phys. Rev. Lett. 93, 202001 (2004).

\bibitem{zzx}C.-H. Chang, C. S. Kim, G.-L. Wang, Phys. Lett. B 623, 218 (2005).

\bibitem{nega}B. Aubert et al., BABAR Collaboration,
hep-ex/0408087; \\
CLEO and FOCUS Collaborations reported their results at ICHEP04,
Beijing 2004.


%%%%%%%%%%%XYZ states Reviews
\bibitem{review8a}E. S. Swanson, Phys. Rept. 429, 243 (2006).
\bibitem{review8b}S. Godfrey, hep-ph/0605152.
\bibitem{review8c}J. L. Rosner, AIP Conf. Proc. 870, 63 (2006),
hep-ph/0606166; hep-ph/0609195, hep-ph/0612332.
\bibitem{review8d}T. Barnes, Int. J. Mod. Phys. A 21, 5583 (2006).
\bibitem{review8e}K. K. Seth, hep-ex/0611035.
\bibitem{review8f}A. Vairo, hep-ph/0611310.

%%%%%%%%%%%%%%%%%%%%%%%%%%%%%%%%%%%%%%%%%%%%%%%%%%%%%%%%%%%%%%%%%
\bibitem{belle8a}S. K. Coi et al., Phys. Rev. Lett. 91, 262001 (2003).%B decay
\bibitem{cdf8a}D. Acosta et al., Phys. Rev. Lett. 93, 072001 (2004).%PPbar
\bibitem{d08}V. M. Abazov et al., Phys. Rev. Lett. 93, 162002 (2003).%ppbar
\bibitem{babar8a}B. Aubert et al., Phys. Rev. D 71, 071103 (2005).%%B decay
\bibitem{babar8b}B. Aubert et al., Phys. Rev. D 73, 011103
(2006).%%B decay
\bibitem{belle8b}K. Abe et al., hep-ex/0505037. %%%%%%%gamma J/psi +omega Jpsi mode
\bibitem{belle8c}K. Abe et al., hep-ex/0505038. %%%%%%%Quantum number
\bibitem{cdf8b}A. Abulencia et al., Phys. Rev. Lett. 96, 102002
(2006).%%%%%%%Dipion spectrum
\bibitem{babar8c}B. Aubert et al., Phys. Rev. D 74,
071101 (2006). %%% gamma Jpsi mode
\bibitem{cdf8c}A. Abulencia et al.,
hep-ex/0612053.%%Quantum number from angular distribution correlation

%%%%%%%%%%%Charmonium
\bibitem{barnes8-qm}T. Barnes, S. Godfrey, Phys. Rev. D 69, 054008 (2004);
T. Barnes, S. Godfrey, E. S. Swanson, Phys. Rev. D 72, 054026
(2005).
\bibitem{eichten8}E. J. Eichten, K. Lane, C. Quigg, Phys. Rev. D
69, 094019 (2004).

%%%%%%%%%Moelcular states
\bibitem{close8}F. E. Close, P. R. Page, Phys. Lett. B 578, 119
(2004).
\bibitem{v8}M. B. Voloshin, Phys. Lett. B 579, 316 (2004).
\bibitem{wong8}C. Y. Wong, Phys. Rev. C 69, 055202 (2004).
\bibitem{swanson8}E. S. Swanson, Phys. Lett. B 588, 189 (2004);
{\bf ibid} B 598, 197 (2004).
\bibitem{tornqvist8}N. A. Tornqvist, Phys. Lett. B 590, 209 (2004).

%%%%%%%%%%%%%a 1++ cusp due to the Dbar-D* threshold
\bibitem{bugg8}D. V. Bugg, Phys. Lett. B 598, 8 (2004).
%%%%%%%%S-wave threshold
\bibitem{swave8}J. L. Rosner, Phys. Rev. D 74, 076006 (2006).

%%%%%%%%%%%%%%%%%%Hybrid
\bibitem{lba8}B. A. Li, Phys. Lett. B 605, 306 (2005).

%%%%%%%%%%%%%%%%%%diquark-Antidiquark tetraquark
\bibitem{miani8}L. Maiani, F. Piccinini, A. D.
Polosa, V. Riquer, Phys. Rev. D 71, 014028 (2005).
\bibitem{richard8}H. Hogaasen, J.M. Richard, P. Sorba, Phys. Rev. D 73, 054013
(2006).
\bibitem{ebert8}D. Ebert, R. N. Faustov, V. O. Galkin, Phys. Lett.
B 634, 214 (2006).
\bibitem{vijande8}N. Barnea, J. Vijande, A. Valcarce, Phys. Rev. D 73,
054004 (2006).
\bibitem{cui8}Y. Cui, X. L. Chen, W. Z. Deng, Shi-Lin Zhu,
High Energy Phys. Nucl. Phys. 31, 7 (2007), hep-ph/0607226. R.D.
\bibitem{narison8}R. D. Matheus, S. Narison, M. Nielsen,
J. M. Richard, Phys. Rev. D 75, 014005 (2007).
\bibitem{tw8}T. W. Chiu et al., Phys. Lett. B 646, 95 (2007);
Phys. Rev. D 73, 111503 (2006), Erratum-ibid.D75:019902,2007.

\bibitem{hy8}K. J. Juge, J. Kuti, and C. J. Morningstar, Phys. Rev. Lett.
82, 4400 (1999);X. Liao and T. Manke, hep-lat/0210030; K. J. Juge,
J. Kuti, and C. Morningstar, Phys. Rev. Lett. 90, 161 601 (2003).

%%%%%%%%%%%%%%%%%%Braaten Production%%%%%%%%%%%%%%%%%%%%%%%%
\bibitem{bra8}E. Braaten, M. Kusunoki, Phys. Rev. D. 71,
074005 (2005).

%%%%%%%%%Belle EXpt on D D pi mode!
\bibitem{belle8d}G. Gokhroo et al., Phys. Rev. Lett. 97, 162002
(2006). %%DDpi mode!

%%%%%%%%%%Charmonium arguments
\bibitem{chao8}C. Meng, Y. J. Gao, K. T Chao, hep-ph/0506222.
\bibitem{suzuki8}M. Suzuki, Phys. Rev. D 72, 114013 (2005).

%%%%%%%%%Final State Interaction
\bibitem{fsi8}X. Liu, B. Zhang, Shi-Lin Zhu, Phys. Lett.
B 645, 185 (2007).
\bibitem{meng8}C. Meng, K.-T. Chao, hep-ph/0703205.

%%%%%%%%%%%%%%%%%%%%%lattice Please check!
\bibitem{liuchuan}Y. Chen et al., CLQCD Collaboration,
hep-lat/0701021.

\bibitem{chao8a}K. T. Chao, private communications.

%%%%%%%%%%%%%%%%%%%%%%%%%%%%%%%%%%%%%%%%%%%%%%%%%%%%%%%%%%%
\bibitem{babar7}B. Aubert et al., BABAR Collaboration,
Phys. Rev. Lett. 95, 142001 (2005).

\bibitem{cleo7}T. E. Coan et al., CLEO Collaboration,
Phys. Rev. Lett. 96, 162003 (2006).

\bibitem{cleo7a}Q. He et al., CLEO Collaboration,
Phys. Rev. D 74, 091104 (2006).

\bibitem{belle7}K. Abe et al., BELLE Collaboration,
hep-ex/0612006.

\bibitem{babar7c}B. Aubert et al., BABAR Collaboration,
Phys. Rev. D 73, 011101 (2006).

\bibitem{mo}X. H. Mo et al., Phys. Lett. B 640, 182 (2006).

\bibitem{liu7}X. Liu, X.-Q. Zeng, X.-Q. Li, Phys. Rev. D 72, 054023 (2005).

\bibitem{yuan7}C.Z. Yuan, P. Wang, X.H. Mo, Phys. Lett. B 634, 399 (2006).

\bibitem{qiao7}C. F. Qiao, Phys. Lett. B 639, 263 (2006).

\bibitem{jlr7}J. L. Rosner, Phys. Rev. D 74, 076006 (2006).

\bibitem{be7}E. van Beveren, G. Rupp, hep-ph/0605317.

\bibitem{miani7}L. Maiani, V. Riquer, F. Piccinini, A. D. Polosa,
Phys. Rev. D 72, 031502 (2005).

\bibitem{4s}F. J. Llanes-Estrada, Phys. Rev. D 72, 031503 (2005).

\bibitem{eichten7}E. J. Eichten, K. Lane, C. Quigg, Phys. Rev.
D 73, 014014 (2006), Erratum-ibid. D 73, 079903 (2006).

\bibitem{rosner7}C. Quigg, J. L. Rosner, Phys. Lett. B 71, 153 (1977).

\bibitem{chao7}Y. B. Ding, K. T. Chao, D. H. Qin, Phys. Rev. D 51, 5064
(1995).

\bibitem{zhu7}Shi-Lin Zhu, Phys. Lett. B 625, 212 (2005).

\bibitem{kou7}E. Kou, O. Pene, Phys. Lett. B 631, 164 (2005).

\bibitem{close7}F. E. Close, P. R. Page, Phys. Lett. B 628, 215 (2005).

\bibitem{tw}T. W. Chiu and T. H. Hsieh, Phys. Rev. D 73, 094510
(2006).

\bibitem{luo}X. Q. Luo, Y. Liu, Phys. Rev. D 74, 034502 (2006);
Erratum-ibid. D74, 039902 (2006).

\bibitem{babar7b}B. Aubert et al., BABAR Collaboration,
hep-ex/0610057.

\bibitem{belle9z}S. Uehara et al., Phys. Rev. Lett. 96, 082003 (2006).

\bibitem{belle9x}K. Abe et al, hep-ex/0507019.

\bibitem{belle9y}S.-K. Choi et al., Phys. Rev. Lett. 94, 182002
(2005).

\end{thebibliography}
\end{document}